\documentclass[apj]{emulateapj}
\usepackage{graphicx,times,natbib,longtable,placeins}

\newcommand{\sersic}{{S\'{e}rsic }}

\def\hst{{\it HST}}
\def\spitzer{{\it Spitzer}}
\def\chandra{{\it Chandra}}

\def\spose#1{\hbox to 0pt{#1\hss}}
\def\lta{\mathrel{\spose{\lower 3pt\hbox{$\mathchar"218$}}
     \raise 2.0pt\hbox{$\mathchar"13C$}}}

\shorttitle{CANDELS GOODS-S Multi-wavelength Catalog}
\shortauthors{Guo et al.}

\begin{document}

\title{CANDELS Multi-wavelength Catalogs: Source Detection and Photometry in the GOODS-South Field}
\author{Yicheng Guo$^{1,2}$, Henry C. Ferguson$^{3}$, Mauro Giavalisco$^{2}$, Guillermo Barro$^{1}$, S. P. Willner $^{4}$, Matthew L. N. Ashby$^{4}$, Tomas Dahlen$^{3}$, Jennifer L. Donley$^{5}$, Sandra M. Faber$^{1}$, Adriano Fontana$^{6}$, Audrey Galametz$^{6}$, Andrea Grazian$^{6}$, Kuang-Han Huang$^{3,7}$, Dale D. Kocevski$^{8}$, Anton M. Koekemoer$^{3}$, David C. Koo$^{1}$, Elizabeth J. McGrath$^{9}$, Michael Peth$^{7}$, Mara Salvato$^{10,11}$, Stijn Wuyts$^{10}$, Marco Castellano$^{6}$, Asantha R. Cooray$^{12}$, Mark E. Dickinson$^{13}$, James S. Dunlop$^{14}$, G. G. Fazio$^{4}$, Jonathan P. Gardner$^{15}$, Eric Gawiser$^{16}$, Norman A. Grogin$^{3}$, Nimish P. Hathi$^{17}$, Li-Ting Hsu$^{10}$, Kyoung-Soo Lee$^{18}$, Ray A. Lucas$^{3}$, Bahram Mobasher$^{19}$, Kirpal Nandra$^{10}$, Jeffery A. Newman$^{20}$, and Arjen van der Wel$^{21}$}
\affil{$^1$ UCO/Lick Observatory, Department of Astronomy and Astrophysics, University of California, Santa Cruz, CA, USA; {\it ycguo@ucolick.org}}
\affil{$^2$ Department of Astronomy, University of Massachusetts, Amherst, MA, USA}
\affil{$^3$ Space Telescope Science Institute, Baltimore, MD, USA}
\affil{$^4$ Harvard-Smithsonian Center for Astrophysics, Cambridge, MA, USA}
\affil{$^5$ Los Alamos National Laboratory, Los Alamos, NM, USA}
\affil{$^6$ INAF - Osservatorio di Roma, I-00040 Monteporzio, Italy}
\affil{$^7$ Department of Physics and Astronomy, The Johns Hopkins University, Baltimore, MD, USA}
\affil{$^8$ Department of Physics and Astronomy, University of Kentucky, Lexington, KY, USA}
\affil{$^9$ Department of Physics and Astronomy, Colby College, Waterville, ME, USA}
\affil{$^{10}$ Max-Planck-Institut fur Extraterrestrische Physik, D-85741 Garching bei Munchen, Germany}
\affil{$^{11}$ Excellence Cluster, Boltzmann Strasse 2, D-85748 Garching bei Munchen, Germany}
\affil{$^{12}$ Department of Physics and Astronomy, University of California, Irvine, CA, USA}
\affil{$^{13}$ National Optical Astronomy Observatory, Tucson, AZ, USA}
\affil{$^{14}$ Institute for Astronomy, University of Edinburgh, Royal Observatory, Edinburgh EH9 3HJ, UK}
\affil{$^{15}$ NASA’s Goddard Space Flight Center, Astrophysics Science Division, Observational Cosmology Laboratory, Greenbelt, MD, USA}
\affil{$^{16}$ Department of Physics and Astronomy, Rutgers University, New Brunswick, NJ, USA }
\affil{$^{17}$ Carnegie Observatories, Pasadena, CA, USA}
\affil{$^{18}$ Department of Physics, Purdue University, West Lafayette, IN, USA}
\affil{$^{19}$ Department of Physics and Astronomy, University of California, Riverside, CA, USA}
\affil{$^{20}$ Department of Physics and Astronomy, University of Pittsburgh, Pittsburgh, PA, USA}
\affil{$^{21}$ Max-Planck Institute for Astronomy, D-69117 Heidelberg, Germany}



\begin{abstract} 
We present a UV-to-mid infrared multi-wavelength catalog in the CANDELS/GOODS-S
field, combining the newly obtained CANDELS \hst/WFC3 F105W, F125W, and F160W
data with existing public data. The catalog is based on source detection in the
WFC3 F160W band. The F160W mosaic includes the data from CANDELS deep and wide
observations as well as previous ERS and HUDF09 programs. The mosaic reaches a
5$\sigma$ limiting depth (within an aperture of radius 0\farcs17) of 27.4,
28.2, and 29.7 AB for CANDELS wide, deep, and HUDF regions, respectively. The
catalog contains 34930 sources with the representative 50\% completeness
reaching 25.9, 26.6, and 28.1 AB in the F160W band for the three regions. In
addition to WFC3 bands, the catalog also includes data from UV (U-band from
both CTIO/MOSAIC and VLT/VIMOS), optical (\hst/ACS F435W, F606W, F775W, F814W,
and F850LP), and infrared (\hst/WFC3 F098M, VLT/ISAAC Ks, VLT/HAWK-I Ks, and
Spitzer/IRAC 3.6, 4.5, 5.8, 8.0 $\mu$m) observations. 
The catalog is validated via stellar colors, comparison with other published
catalogs, zeropoint offsets determined from the best-fit templates of the
spectral energy distribution of spectroscopically observed objects, and the
accuracy of photometric redshifts. 
The catalog is able to detect unreddened star-forming (passive) galaxies with
stellar mass of ${\rm 10^{10}M_\odot}$ at a 50\% completeness level to
z$\sim$3.4 (2.8), 4.6 (3.2), and 7.0 (4.2) in the three regions. 
As an example of application, the catalog is used to select both star-forming
and passive galaxies at z$\sim$2--4 via the Balmer break. It is also used
to study the color--magnitude diagram of galaxies at 0$<$z$<$4.
\end{abstract}

\section{Introduction}
\label{intro}

Over the past decade, the Hubble Space Telescope (\hst) and modern giant
telescopes have opened a new era in observational cosmology. Galaxies are now
routinely found in the very early universe, and their evolution can be followed
from its early stages to the present. Particularly, deep multi-wavelength
imaging surveys have been established as a powerful and efficient tool to push
observations up to high redshift and down to faint populations of galaxies.
These modern surveys have revealed a complex interplay between galaxy mergers,
star formation, and black holes over cosmic time and have led to new insights
into the physical processes that drive galaxy formation and evolution.

Of the entire cosmic time, two epochs are particularly of great interest:
z$\sim$2 (Cosmic High Noon)  and z$\sim$8 (Cosmic Dawn). In the former, the
cosmic star formation history reached its peak
\citep[e.g.,][]{hopkinsa04,hopkinsa06,bouwens09}, and galaxies quickly
assembled their stars \citep[e.g.,][]{daddi07a,reddy08,reddy09} as well as
began to differentiate into morphological sequence
\citep[e.g.,][]{cassata08,kriek09,whitaker11,wangtao12}. In the latter,
detection and physical properties of galaxies place essential constraints on
the reionization
\citep[e.g.,][]{fontana10,oesch10,grazian11,bouwens12,finkelstein12a}, the
accretion history of galaxies at early times
\citep[e.g.,][]{bouwens10a,finkelstein10,mclure10,trenti11,yanhj12} as well as
the formation of metals and dust in the intergalactic medium
\citep[e.g.,][]{bouwens10b,dunlop12,finkelstein12b}. Because two important
spectral features, namely, the Balmer break at z$\sim$2 and the Lyman break at
z$\sim$8, are redshifted to near-infrared (NIR) bands at the two epochs, a NIR
survey with deep sensitivity and high spatial resolution is required for robust
investigation of galaxies during these epochs.  Incorporating these deep NIR
data into a multi-wavelength catalog would significantly improve measurements
of galaxy properties at z$>$2.
 
The Cosmic Assembly Near-infrared Deep Extragalactic Legacy Survey
\cite[CANDELS,][]{candelsoverview,candelshst} is designed to document galaxy
formation and evolution over the redshift range of z=1.5--8. The core of
CANDELS is to use the revolutionary near-infrared \hst/WFC3 camera, installed
on Hubble in May 2009, to obtain deep imaging of faint and faraway objects.
CANDELS also uses \hst/ACS to carry out parallel observations, aiming to
providing unprecedented panchromatic coverage of galaxies from optical
wavelengths to the near-IR.  Covering approximately 800 ${\rm arcmin^2}$,
CANDELS will image over 250,000 distant galaxies within five popular sky
regions which possess rich existing data from multiple telescopes and
instruments: GOODS-S, GOODS-N, UDS, EGS, and COSMOS. The strategy of five
widely separated fields mitigates cosmic variance and yields statistically
robust and complete samples of galaxies down to a stellar mass of ${\rm 10^9
M_\odot}$ to z$\sim$2, and around the knee of the ultraviolet luminosity
function of galaxies to z$\sim$8. It will also find and measure Type Ia
supernovae at z$>$1.5 to test their accuracy as standard candles for cosmology. 

The GOODS-S field, centered at $\alpha$(J2000) = 03h32m30s and $\delta$(J2000)
= -27d48m20s and located within the Chandra Deep Field South
\citep[CDF-S,][]{giacconi02}, is a sky region of about 170 arcmin${\rm ^2}$ which has
been targeted for some of the deepest observations ever taken by NASA's Great
Observatories: \hst, \spitzer, and \chandra\ as well as by other world-class
telescopes. The field has been imaged in the optical wavelength with \hst/ACS
in F435W, F606W, F775W, and F850LP bands as part of the \hst\ Treasury Program:
the Great Observatories Origins Deep Survey
\citep[GOODS,][]{giavalisco04goods}; in the mid-IR (3.6--24 $\mu$m) wavelength
with \spitzer\ as part of the GOODS \spitzer\ Legacy Program (PI: M.
Dickinson); and in the X-ray with the deepest \chandra\ sky survey ever taken,
with a total integration time of 4 Ms \citep{xue11}. The field also has
very deep far-IR to submillimeter data from {\it Herschel} Space
Observatory from the PEP survey \citep{lutz11} and the HerMes survey
\citep{oliver12}, deep submillimeter (1.1 mm) observation with AzTEC/ASTE
\citep{scott10}, and radio imaging from the VLA \citep{kellermann08}, ATCA
\citep{norris06, zinn12}, and VLBA \citep{middelberg11}. Furthermore, the field
has been the subject of numerous spectroscopic surveys
\citep[e.g.,][]{lefevre04,szokoly04,mignoli05,cimatti08,vanzella08,popesso09,balestra10,silverman10}.
These rich and deep existing data, covering the full wavelength range, make
GOODS-S one of the best fields to study galaxy formation and evolution.

This paper presents a multi-wavelength catalog of GOODS-S based on CANDELS WFC3
F160W detection, combining both CANDELS data and available public imaging data
from UV to mid-IR wavelength. The catalog includes U-band images obtained with
the CTIO Blanco telescope and with the Visible Multi-Object Spectrograph
(VIMOS) on the Very Large Telescope (VLT), \hst/ACS images in F435W, F606W,
F775W, F814W, and F850LP bands, \hst/WFC3 images in F098M, F105W, F125W, and
F160W bands, Ks-band image with the Infrared Spectrometer and Array Camera
(ISAAC) and the High Acuity Wide field K-band Imager (HAWK-I) on the VLT, as
well as \spitzer\ images at 3.5, 4.5, 5.8, and 8.0 $\mu$m. 

The paper is organized as follows: Sec. 2 briefly summarizes the datasets
included in our catalog. Sec. 3 discusses source detection in the CANDELS F160W
image and photometry measurement on \hst\ images. Sec. 4 discusses photometry
measurements on low resolution images. Sec. 5 evaluates the quality of our
photometry catalog. Sec. 6 presents a few simple applications of our catalog to
study galaxy formation and evolution. The summary is given in Sec. 7.

All magnitudes in the paper are on the AB scale \citep{oke74} unless otherwise
noted. We adopt a flat ${\rm \Lambda CDM}$ cosmology with $\Omega_m=0.3$,
$\Omega_{\Lambda}=0.7$ and use the Hubble constant in terms of $h\equiv H_0/100
{\rm km~s^{-1}~Mpc^{-1}} = 0.70$.  

The CANDELS GOODS-S multi-wavelength catalog and its associated files ---
all SExtractor photometry of the \hst\ bands as well as the system throughput
of all included filters --- are made publicly available on the CANDELS
Website\footnote[1]{http://candels.ucolick.org}, in the Mikulski Archive for
Space Telescopes (MAST)\footnote[2]{http://archive.stsci.edu}, via the
on-line version of the article, and the Centre de Donnees astronomiques de
Strasbourg (CDS). They are also available in the Rainbow
Database\footnote[3]{US:
https://arcoiris.ucolick.org/Rainbow\_navigator\_public, and Europe:
https://rainbowx.fis.ucm.es/Rainbow\_ navigator\_public.} \citep{pg08,
barro11a}, which features a query menu that allows users to search for
individual galaxies, create subsets of the complete sample based on different
criteria, and inspect cutouts of the galaxies in any of the available bands. It
also includes a cross-matching tool to compare against user uploaded catalogs.

\section{Data}
\label{data}

\begin{figure*}[htbp]
\center{\includegraphics[scale=0.45, angle=0]{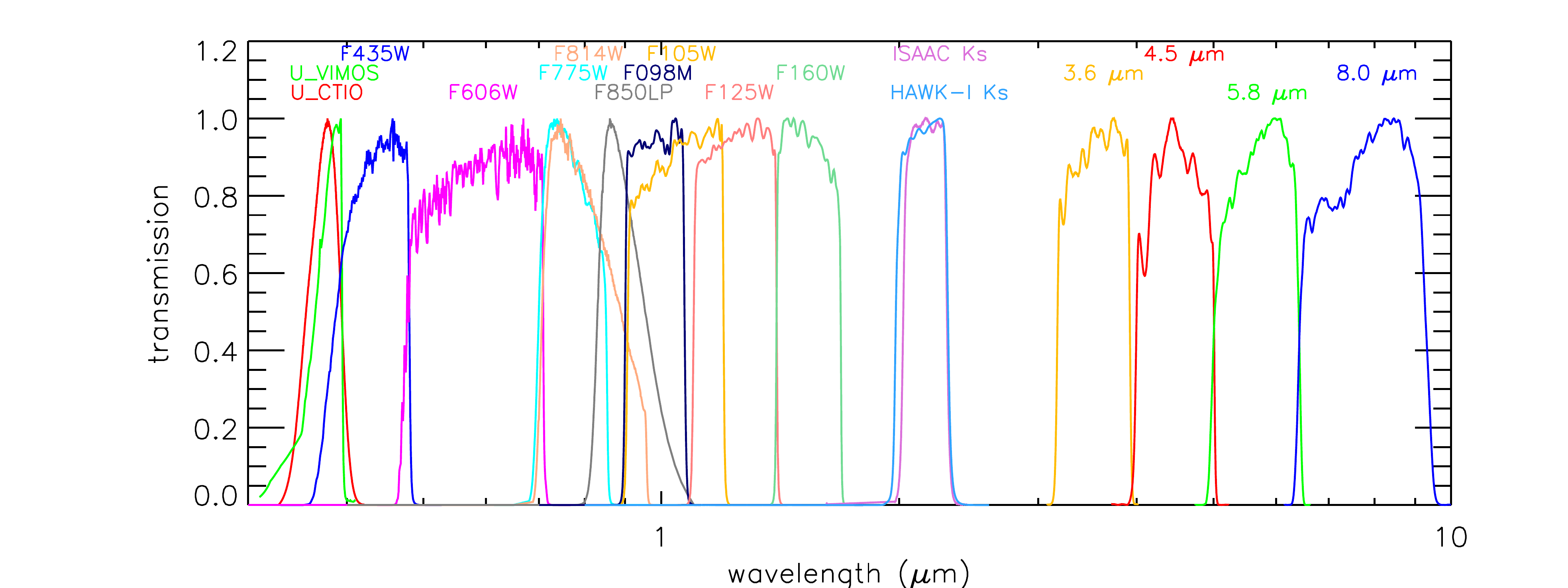}}
\caption[]{Transmission of all filters used in our multi-wavelength catalog 
in the CANDELS GOODS-S field. All transmissions are normalized to have a maximum 
value of unity.
\label{fig:filter}}
\vspace{-0.2cm}
\end{figure*}

\begin{figure}[t]
\center{\includegraphics[scale=0.45, angle=0]{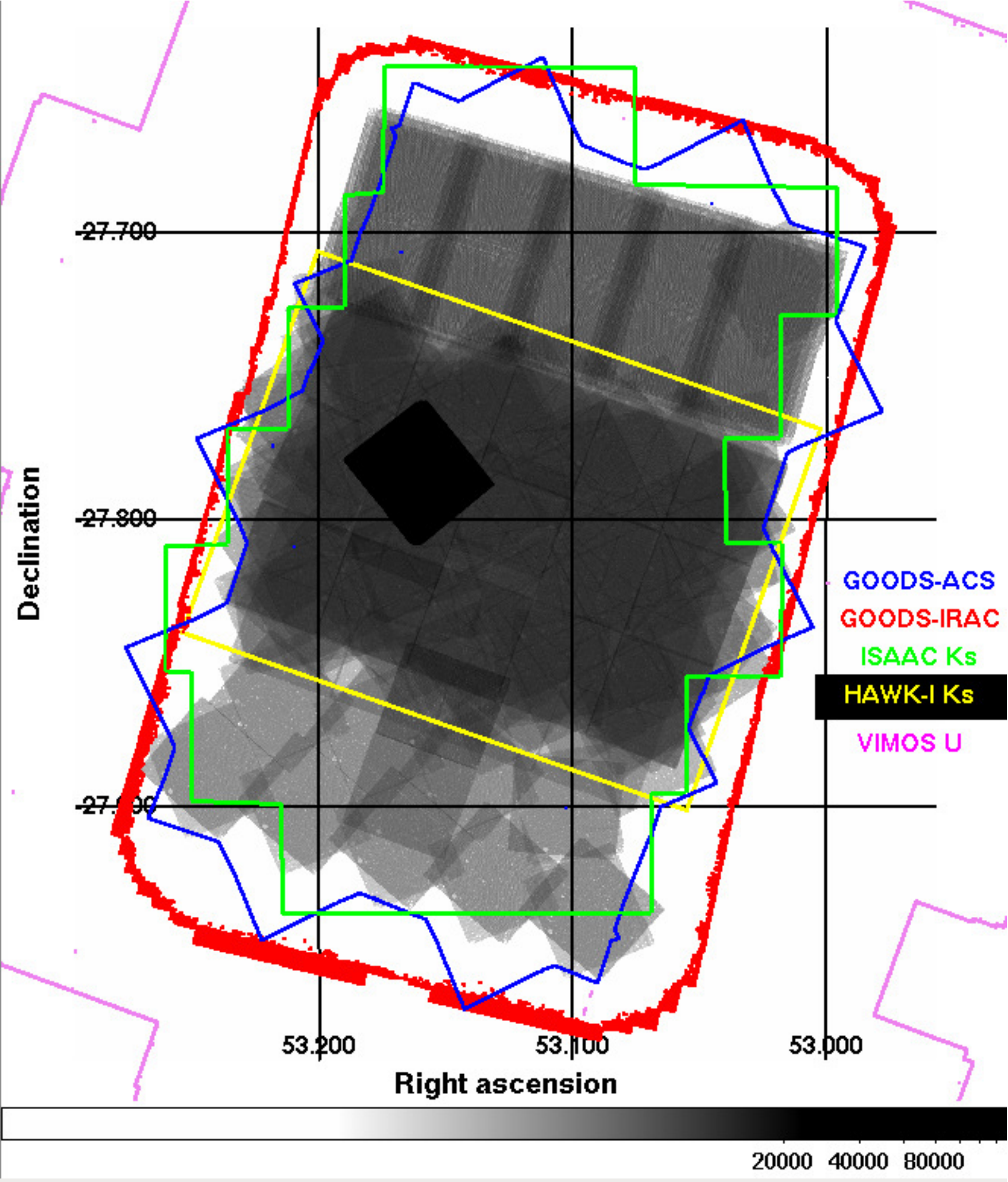}}
\caption[]{Sky coverage of data sets used in our catalog. The gray scale 
image shows the
exposure time of our max-depth F160 mosaic, which includes the CANDELS wide and
deep region, ERS, and HUDF09. Coverage of ancillary data from UV to MIR are
also shown: VLT/VIMOS U (magenta), GOODS \hst/ACS (blue), VLT/ISAAC Ks (green),
VLT/HAWK-I Ks (yellow), and GOODS \spitzer/IRAC (red). The entire field is 
covered by both SEDS \spitzer/IRAC and VLT/VIMOS U.
\label{fig:map}}
\vspace{-0.2cm}
\end{figure}

Table \ref{tb:band} summarizes the multi-wavelength datasets included in our
catalog. The table lists the effective wavelength, the full width at
half-maximum (FWHM) of the filter, the FWHM of the point spread function (PSF),
and the limiting magnitude of each band in our catalog.  For each band, we
calculate the 5$\sigma$ limiting magnitude with an aperture whose radius is
equal to the FWHM of the band. No aperture correction is applied to the
limiting magnitude.  Transmission curves of all filters used in our catalog are
plotted in Figure \ref{fig:filter}. The sky coverage of observations used in
our catalog is shown in Figure \ref{fig:map}. 

\begin{table*}[htbp]
\begin{center}
\caption{Summary of Data Included in the CANDELS GOODS-S Catalog \label{tb:band}}
\begin{tabular}{cccccc}
\hline\hline
Instrument & Filter & Effective     & Filter  & PSF       & 5$\sigma$ Limiting  \\
           &        & Wavelength\footnote[1]{Effective wavelength is calculated as: $\lambda_{eff} = \sqrt{(\int S(\lambda) \lambda d\lambda) / (\int S(\lambda) \lambda^{-1} d\lambda)}$ \citep{tokunaga05}}  & FWHM    & FWHM      & Depth       \\
           &        &  (\AA)      & (\AA)   & (arcsec)  & (AB)\footnote[2]{Aperture magnitude at 5$\sigma$ within ${\rm r_{ap}=1}$ FWHM of the PSF. Point source magnitude will be brighter than this by the aperture correction assuming the profile of the PSF. Aperture corrections for extended sources will depend on their surface-brightness profiles.} \\
\hline
Blanco/MOSAIC II & U\_CTIO & 3734 & 387 & 1.37 & 26.63 \\
VLT/VIMOS & U\_VIMOS & 3722 & 297 & 0.80 & 27.97 \\
\hst/ACS & F435W & 4317 & 920 & 0.08 & 28.95 / 30.55\footnote[3]{The limiting 
depths of ACS bands are measured within apertures with a fixed radius of 0\farcs09. Each band has two measurements: one for GOODS-S V2.0 and the other for HUDF.} \\
        & F606W & 5918 & 2324 & 0.08 & 29.35 / 31.05$^c$ \\
        & F775W & 7693 & 1511 & 0.08 & 28.55 / 30.85$^c$ \\
        & F814W & 8047 & 1826 & 0.09 & 28.84 \\
        & F850LP & 9055 & 1236 & 0.09 & 28.55 / 30.25$^c$ \\
\hst/WFC3 & F098M & 9851 & 1696 & 0.13 & 28.77 \\
         & F105W & 10550 & 2916 & 0.15 & 27.45 / 28.45 / 29.45\footnote[4]{The limiting depths of WFC3 bands are measured within apertures with a fixed radius of 0\farcs17. Each band has three measurements: for CANDELS-wide, CANDELS-deep, and HUDF09, respectively.}  \\
         & F125W & 12486 & 3005 & 0.16 & 27.66 / 28.34 / 29.78$^d$ \\
         & F160W & 15370 & 2874 & 0.17 & 27.36 / 28.16 / 29.74$^d$ \\
VLT/ISAAC & Ks   & 21605 & 2746 & 0.48$^{+0.1}_{-0.1} $\footnote[5]{PSF and depth vary among ISAAC tiles. We show the median and the 15 and 85 percentiles of all available tiles.} & 25.09$^{+0.60}_{-0.32}$ $^e$ \\
VLT/HAWK-I & Ks   & 21463 & 3250 & $\sim$0.4 & 26.45 \\
Spitzer/IRAC & 3.6$\mu$m & 35508 & 7432 & 1.66  &  26.52\footnote[6]{This is the depth of the SEDS GOODS-S region measured through apertures. For the SEDS non-GOODS region, the 5$\sigma$ depth is $\sim$25.4 AB mag measured through photometry of artificial sources. \citet{ashby13seds} give details of the depth of SEDS depths.}  \\
             & 4.5$\mu$m & 44960 & 10097 & 1.72  & 26.25$^f$ \\
             & 5.8$\mu$m & 57245 & 13912 & 1.88  & 23.75\footnote[7]{This is the average depth of the two GOODS Spitzer epochs. The overlapped region, which is in the CANDELS-deep region, is deeper by about 0.35 mag.} \\
             & 8.0$\mu$m & 78840 & 28312 & 1.98  & 23.72$^g$ \\
\hline
\end{tabular}
\end{center}
\end{table*}

\subsection{HST/WFC3 Imaging}
\label{data:wfc3}

Several programs have carried out observations with \hst/WFC3 IR channels in
GOODS-S, including CANDELS, the \hst/WFC3 Early Release Science
\citep[ERS,][]{windhorst11ers}, and HUDF09 \citep{bouwens10a}. To maximize the
scientific merit of our multi-wavelength catalog, we combine all published
\hst/WFC3 imaging within GOODS-S prior to June, 2012 to make a ``max-depth''
co-added mosaic of \hst/WFC3 images. These co-added images provide each source
the deepest NIR observation to date to ensure the least uncertainty when these
bands are used for measuring photometric redshifts and stellar masses. A
potential drawback is the inhomogeneity of the depth across the whole GOODS-S
field, which may bring complexity for measuring the completeness of the
catalog. However, even within each sub-region of the field, for example the
CANDELS wide region, the actual point-by-point exposure times and sensitivities
vary considerably. Therefore, the inhomogeneity issue exists anyway. 

CANDELS observed GOODS-S with the \hst/WFC3 F105W, F125W, and F160W filters using
two strategies: deep and wide. The deep region covers the middle one-third of
the GOODS-S ACS region with an area of $\sim$55 arcmin${\rm ^2}$ by 3, 4, and 6
orbits with the F105W, F125W, and F160W filters, respectively. The wide region
covers the southern one-third of the GOODS-S ACS region and has approximately
1-orbit exposure for all three bands. \citet{candelsoverview} and
\citet{candelshst} give details of CANDELS \hst/WFC3 observations, survey
design, and data reduction. The GOODS-S region was also observed by CANDELS with
the \hst/ACS F814W filter in parallel.

The northern one-third of the GOODS-S region was observed by \hst/WFC3 ERS
\citep{windhorst11ers} with the F098M, F125W, and F160W filters. ERS covered
this region with 2 orbits in each of the three bands.

\begin{figure}[htbp]
\vspace*{-0.7cm}
\center{
\hspace*{-1.0cm}
\includegraphics[scale=0.33, angle=0]{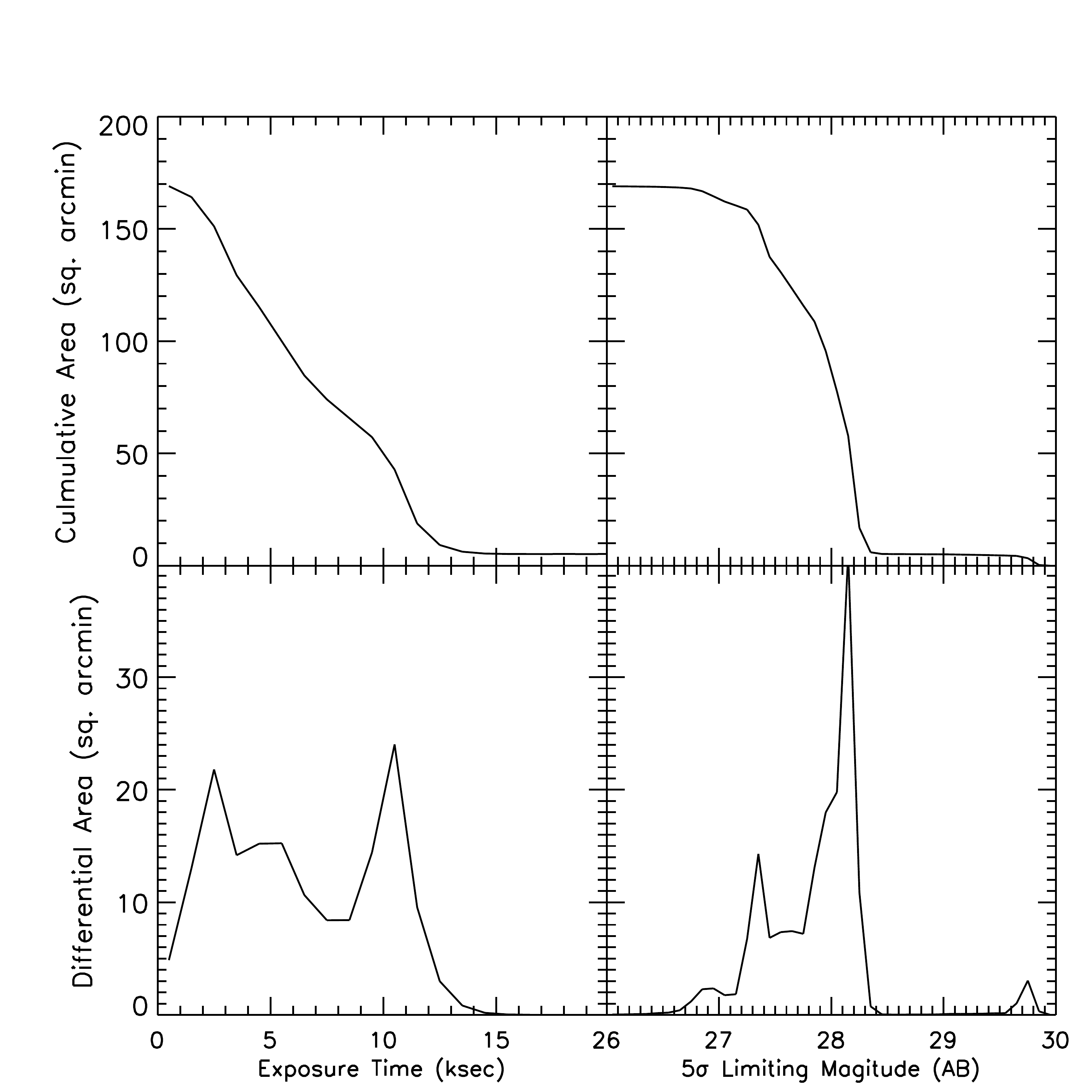}}
\caption[]{Distributions of exposure time and limiting magnitude of our
max-depth F160W mosaic used as the detection image of our catalog. The {\it
left} column shows the cumulative (upper panel) and differential (lower)
distributions of the exposure time, while the {\it right} column shows the same
distributions of 5$\sigma$ limiting magnitude of the image. HUDF09 is not
shown in the left column as its exposure time ($\sim$155 ks) is far off-axis.
\label{fig:limitmag}}
\vspace{-0.0cm}
\end{figure}

An area of about 4.6 arcmin${\rm ^2}$ in GOODS-S, the Hubble Ultra Deep Field
(HUDF), is covered by ultra-deep \hst/WFC3 imaging from G. Illingworth's HUDF09
program \citep[GO 11563,][]{bouwens10a}. HUDF09 observed the field for 24, 34,
and 53 orbits in the F105W, F125W, and F160W bands. The CANDELS deep/wide
region and ERS, together with the Hubble Ultra Deep Fields, create a
three-tiered “wedding-cake” approach that has proven efficient for
extragalactic surveys. We carry out our own data reduction on both ERS and HUDF
images and drizzled them to 0\farcs06/pixel to match our CANDELS pixel scale
\citep[see][for details]{candelshst}. The distributions of exposure time and
limiting magnitude of our max-depth F160W mosaic are shown in Figure
\ref{fig:limitmag}.

\subsection{HST/ACS Imaging}
\label{data:acs}

The \hst/ACS images used in our catalog are the version v3.0 of the mosaicked
images from the GOODS \hst/ACS Treasury Program. They consist of data acquired
prior to the \hst\ Servicing Mission 4, including mainly data of the original
GOODS \hst/ACS program in \hst\ Cycle 11 \citep[GO 9425 and 9583;
see][]{giavalisco04goods} and additional data acquired on the GOODS fields
during the search for high redshift Type Ia supernovae carried out during
Cycles 12 and 13 \citep[Program ID 9727, P.I. Saul Perlmutter, and 9728, 10339,
10340, P.I. Adam Riess; see, e.g.,][]{riess07}. The GOODS-S field was observed
in ACS B, V, i, and z-band with a total exposure time of 7200, 5450, 7028, and
18232 seconds. 

\subsection{Ground-based Imaging}
\label{data:gb}

GOODS-S is covered by a large number of ground-based images. Combined into our
catalog is imaging of CTIO U-band, VLT/VIMOS U-band, VLT/ISAAC Ks-band, and
VLT/HAWK-I Ks-band.

The CDF-S/GOODS field was observed by the MOSAIC II imager on the CTIO 4m
Blanco telescope to obtain deep U-band observations in September, 2001. The
observations were taken through CTIO filter c6021, the SDSS
u$^\prime$ filter\footnote[1]{A red leak is found in the filter, between 7000
and 11000 \AA. The red leak affects the photometry of objects with red colors.
\citet{smith02} found that the red leak makes objects with red color
(u$^\prime$-i$^\prime >$5 mag) brighter in the u$^\prime$-band than they should
be.}. This filter was chosen in order to minimize bandpass overlap with the
standard B-band and the ACS F435W filter in hope of providing better
photometric redshift and Lyman break constraints on z$\sim$3 galaxies. The
final image has a total integration time of about 17.5 hours. 

Another U-band survey in GOODS-S was carried out using the VIMOS instrument
mounted at the Melipal Unit Telescope of the VLT at ESO's Cerro Paranal
Observatory, Chile. This large program of ESO (168.A-0485, PI C. Cesarsky) was
obtained in service mode observations in UT3 between August 2004 and fall 2006,
with a total time allocation of 40 hours. \citet{nonino09} give details of the
U-band imaging.

In the ground-based NIR, imaging observations of the CDF-S were carried out in J, H, Ks bands using the ISAAC instrument
mounted at the Antu Unit Telescope of the VLT. Data were obtained as part of the ESO Large Programme
168.A-0485 (PI: C. Cesarsky) as well as ESO Programmes 64.O-0643, 66.A-0572 and
68.A-0544 (PI: E. Giallongo) with a total allocation time of $\sim$500 hours
from October 1999 to January 2007. The data cover 172.4, 159.6, and 173.1
arcmin${\rm ^2}$ of the GOODS/CDF-S region in J, H, and Ks, respectively.
Because the CANDELS \hst/WFC3 F125W and F160W bands surpass the ground-based J-
and H-bands in both spatial resolution and sensitivity, we use only the Ks data
from the GOODS VLT/ISAAC program. \citet{retzlaff10} give details of the ISAAC
Ks-band imaging.

The CANDELS/GOODS-S field was also observed in the NIR as part of the
ongoing HAWK-I UDS and GOODS-S survey (HUGS; VLT large program ID 186.A-0898,
PI: A.  Fontana; Fontana et al. in prep.) using the High Acuity Wide field
K-band Imager (HAWK-I) on VLT. Included in this paper are the two deep HAWK-I
pointings in the CANDELS deep region. 
Each pointing was imaged in the Ks-band for $\sim$31 hours. Fontana et al.
(in preparation) will give more details of the HUGS survey and data.

\subsection{Spitzer/IRAC Imaging}
\label{data:irac}

GOODS-S was observed by \spitzer/IRAC \citep{fazio04} with four channels (3.6,
4.5, 5.8, and 8.0 $\mu$m) in two epochs with a separation of six months (February 2004
and August 2004) by the GOODS Spitzer Legacy project (PI: M. Dickinson). Each
epoch contained two pointings, each with total extent approximately 10 arcmin
on a side. IRAC observed simultaneously in all four channels, with channels 1
and 3 (3.6 and 5.8 microns) covering one pointing on the sky and channels 2 and
4 (4.5 and 8.0 microns) covering another pointing. After two epochs, GOODS-S
has complete coverage in all four IRAC channels with an overlap strip in the
middle receiving twice the exposure time of the rest of the field. The exposure
time per channel per sky pointing was approximately 25 hours per epoch and
doubled in the overlap strip. We include the GOODS 5.8 and 8.0 $\mu$m imaging
in our catalog.

The {\sl Spitzer}/IRAC 3.6 and 4.5\,$\mu$m images used here were drawn from the
{\sl Spitzer} Extended Deep Survey \citep[SEDS; PI G. Fazio;][]{ashby13seds}.
The SEDS ECDFS observations were acquired during three separate visits made
during the warm {\sl Spitzer} mission. Each SEDS visit covered a region
surrounding the existing IRAC exposures from the GOODS program, and the SEDS
mosaics incorporate the pre-existing cryogenic observations. The SEDS mosaics
were pixellated to 0\farcs6 and were constructed with a tangent-plane
projection designed to match that of the CANDELS \hst/WFC3 mosaics.  
Total 3$\sigma$ depths in the SEDS-only portions of the field are $\sim$26 AB
mag measured through photometry of artificial sources. Catalogs and source
counts in the 3.6 and 4.5\,$\mu$m mosaics are fully described by
\citet{ashby13seds}.

\section{Photometry of \hst\ Images}
\label{hstphoto}

\subsection{Source Detection in Max-depth F160W Image}
\label{hstphoto:source}

The scientific goals of CANDELS extend from studying the morphology of galaxies at
the cosmic high-noon (z$\sim$2) to searching for faint galaxies at the cosmic
dawn (z$>$7). Therefore, our source detection strategy must be efficient in
detecting both bright/large and faint/small sources.  It is, however, almost
impossible to design a single set of SExtractor parameters to achieve this
goal. A mild detection and deblending threshold avoids breaking large/bright
sources into small pieces but misses a large fraction of faint sources. On the
other hand, an aggressive detection threshold able to detect very faint sources
over-deblends bright/large galaxies. To ensure that our catalog provides
reliable detection for both the bright/large and faint/small sources, we
adopt a two-mode detection strategy, which has been shown to be successful and
efficient in other deep sky surveys, e.g., GEMS \citep{rix04} and STAGES
\citep{gray09}.

We run our modified SExtractor \citep[see][]{galametz13uds} on our
co-added max-depth version of \hst/WFC3 F160W image to detect sources in two
modes: ``cold'' and ``hot''. In the cold mode, the SExtractor configuration is
designed to detect bright/large sources without over-deblending them. In the
hot mode, the SExtractor configuration is pushed to detect faint sources toward
the limiting depth of the image. In this mode, bright/large sources may be
dissolved, and some of their sub-structures are treated as independent
sources. We then follow the strategy of GALAPAGOS \citep{barden12} to merge the
two modes: all cold sources are kept into the final detection catalog, but hot
sources within the Kron radius of any cold sources are treated as
sub-structures of the cold sources and excluded from the final catalog. Only
the hot sources that are not within the Kron radius of any cold sources are
included in the final catalog. We refer readers to GALAPAGOS \citep{barden12}
and \citet{galametz13uds} for details of this two-mode strategy and
its application on CANDELS images. In total, we detect 34930 sources from our
max-depth F160W mosaic. Among them, 26835 sources are detected by the cold mode
and 8095 sources by the hot mode. The key SExtractor parameters used in our
source detection are listed in Table \ref{tb:sex}.

The inhomogeneous depth across our detection image brings a new issue on the
detection and deblending threshold. Ideally, in order to construct a catalog
whose completeness is similar over regions with various depths, one would
use a distinct configuration for each region or use a fixed surface brightness
threshold instead of signal-to-noise ratio (S/N) to detect sources. In our
catalog, however, we use the same S/N threshold for detecting sources in all
regions. The choice is based on two facts.  First, the purpose of our max-depth
catalog is to push our detection ability in each region to its limit.
Therefore, it is the limit on the faint end instead of a homogeneous
completeness that we are interested in. Second, even within one region, the
depth varies from point to point due to the observation pattern. For example, in
the ``wide'' region, the exposure time of points observed by more than one tile
could be 3--4 times higher than that of points observed only once. Because the
inhomogeneity exists not only between different regions but also within each
region, it would be impossible to design a distinct detection configuration for
each region. For readers who are interested in completeness, the catalog
provides the exposure time (or depth) of each source, which can be used to
construct a sample that is complete to the same depth over all regions.

\begin{table}[htbp]
\begin{center}
\caption{SExtractor Parameters in Cold and Hot Modes \label{tb:sex}}
\begin{tabular}{ccc}
\hline\hline
        &   Cold Mode  & Hot Mode \\
\hline
DETECT\_MINAREA &  5.0   & 10.0 \\
DETECT\_THRESH  &  0.75  & 0.7 \\
ANALYSIS\_THRESH & 5.0   & 0.8 \\
FILTER\_NAME  &   tophat\_9.0\_9x9.conv & gauss\_4.0\_7x7.conv \\
DEBLEND\_NTHRESH & 16 & 64 \\
DEBLEND\_MINCONT & 0.0001 & 0.001 \\
BACK\_SIZE    &   256    &  128 \\
BACK\_FILTERSIZE  & 9 & 5 \\
BACKPHOTO\_THICK & 100  & 48 \\
MEMORY\_OBJSTACK & 4000  & 4000 \\       
MEMORY\_PIXSTACK & 400000 & 400000 \\
MEMORY\_BUFSIZE  & 5000 & 5000 \\
\hline
\end{tabular}
\end{center}
\end{table}

\begin{figure*}[htbp]
\center{
\hspace*{-0.7cm} 
\includegraphics[scale=0.48, angle=0]{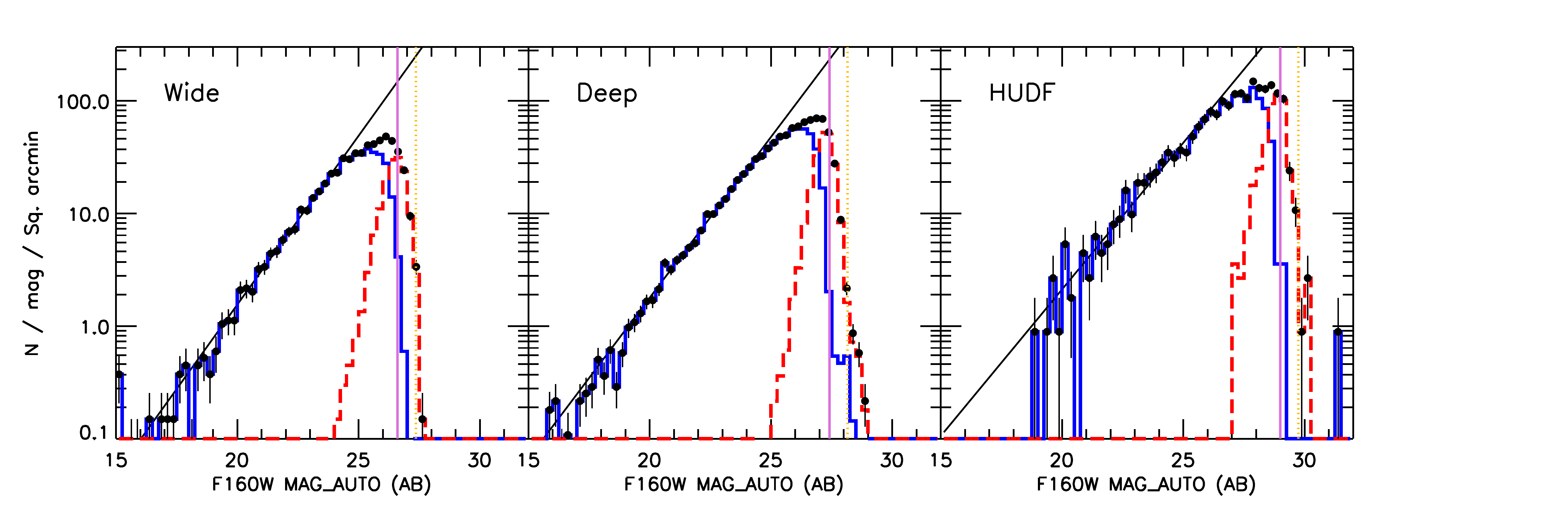}}
\caption[]{Differential number density of objects detected in our
multi-wavelength catalog for CANDELS wide, deep (including ERS) and HUDF
regions. Black points with error bars stand for the differential number density
and its Poisson uncertainty of all detected objects in each region. Solid blue
and dashed red histograms show the contributions of sources detected in the
cold and hot mode, respectively. The solid black line in each panel is the best
power-law fit to the differential number density of all detected sources in the
region in the magnitude range in which the sample is believed to be complete
(see the text for details of the fitting ranges). In each panel, the 
vertical dotted yellow line shows the 5$\sigma$ limiting depth of the region 
without aperture correction (i.e., the value in Table \ref{tb:band}), 
while the vertical solid purple line shows the depth after point source aperture correction.
\label{fig:ndensity}}
\vspace{-0.2cm}
\end{figure*}

The differential number densities of sources detected in the three (wide, deep
+ ERS, and HUDF) regions are shown in Figure \ref{fig:ndensity}. We also
separate the contributions of sources detected in the cold and hot modes. In
all regions, the distribution peak of the hot-detected sources is almost 1--1.5
magnitude fainter than that of the cold-detected ones. The hot mode
significantly improves our detection ability toward the limiting depth of our
images. To make a fair comparison between the total magnitude (MAG\_AUTO) of
detected sources and the limiting depth, we correct the 5$\sigma$ limiting
depths of the F160W band in Table \ref{tb:band}, which is measured within an
aperture, to include the light outside the aperture by assuming the light
profile of the F160W PSF. The peak of the hot detected sources matches the
corrected depth (purple line) very well in all regions. Compared to the cold
mode, the hot mode boosts the source number density at the limiting depth
by a factor of $\sim$20.

\begin{figure}[htbp]
\vspace*{-2.5cm}
\center{
\hspace*{-0.7cm}
\includegraphics[scale=0.45, angle=0]{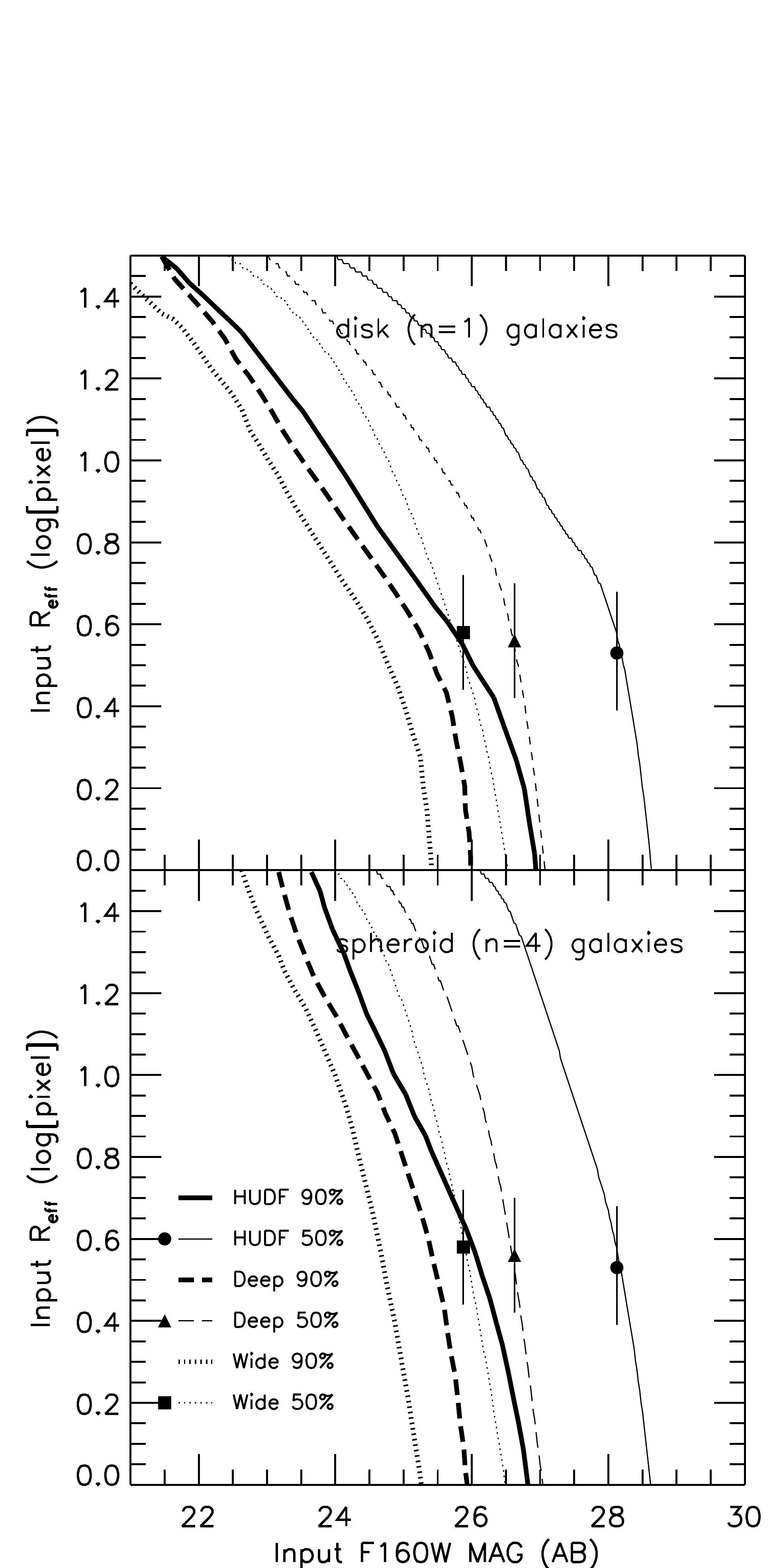}}
\caption[]{Completeness estimated from detecting fake sources
from our F160W mosaics. The light profile of fake sources is assumed to be
an exponential disk (\sersic index n=1, top panel) or a de Vaucouleurs profile
(n=4, bottom panel). In each case, the constant curves of 50\% and 90\%
completeness of each region are plotted in the input size--magnitude diagram,
as the labels show. The 50\% completeness estimated through the best-fit power law
to the actual differential number density (Figure \ref{fig:ndensity}) is also
plotted (symbols with error bars) as a reference. The symbols are placed at the
median of the size (half-light radius measured by SExtractor) distribution of
each region, and their error bars stand for the 20$th$ and 80$th$ percentile of the
size distribution. 
\label{fig:complete}}
\vspace{-0.2cm}
\end{figure}

The differential number density provides a rough way to estimate the
completeness of our catalog in different regions. We fit a power law to the
differential number density of each region. The catalog probably becomes
incomplete when the actual differential number density significantly deviates
from the best-fit power law at the faint end.  We only fit the power law in the
magnitude range in which the catalog is believed to be complete as well as has
good number statistics: 20--24 mag, 20--24 mag, and 21--26 mag for the wide,
deep, and HUDF regions, respectively. 
The actual differential number density becomes lower than the best-fit power
law by a factor of two at 25.9, 26.6, and 28.1 mag for the three regions.
Therefore, our catalog is $\sim$50\% complete at these magnitudes for the three
regions. 

This completeness is only a rough estimate. It depends on the assumption that
our catalog is complete in the fitting range. What is more, it assumes that the
completeness only depends on the flux of sources. It also assumes the
counts are a power law over the magnitude range that was fit, and that this
power law extends to fainter magnitudes. In fact, the completeness depends on
the flux, size, and even morphology of sources. The completeness estimated here
is just an overall value for objects with mixed types and should be used with
caution when dealing with a specific type of object.

In order to accurately estimate the completeness of sources with various
fluxes, sizes, and light profiles, we carry out Monte Carlo simulations of
detecting fake sources with the same detection strategy used in our catalog for
each region. The fake galaxies populate the magnitude range of 20 mag $<$ F160W $<$
30 mag and the range of galaxy half-light radii from 0.1 pixel to 30 pixels.
The input galaxies are spheroids with a de Vaucouleurs surface brightness
profile (\sersic index $n=4$) and disks with an exponential profile ($n=1$).
The simulated galaxies are convolved with the F160W PSF and inserted into our
F160W mosaic with additional Poisson noise. We use the SExtractor
parameters of both our cold and hot modes to recover the fake sources. The
resulting completeness limits in a plane of magnitude and half-light radius are
shown in Figure \ref{fig:complete}.

In this figure, we over-plot the power-law estimated 50\% completeness at the
median size (the half-light radius measured by SExtractor) of sources in each
region. These values are in good agreement with those measured by the simulation
for sources with the input size similar to the median of the actual size
distribution. The 20$th$ and 80$th$ percentiles of the actual size
distributions span a fairly small range of about 0.3 dex, within which the 50\%
completeness measured through the simulation changes mildly for each region.
Therefore, the 50\% completeness estimated from the power law can be used as a
representative limit for the majority of sources in our catalog. For sources
with much larger sizes, however, their completeness is much worse than this
representative value. The completeness also depends on the morphology of
sources. Sources with higher light concentration ($n=4$) are complete to
fainter magnitude than those with lower concentration ($n=1$) at the same
radius.

\subsection{Photometry of Other \hst\ Images}
\label{hstphoto:others}

Photometry of other \hst\ bands, namely, ACS F435W, F606W, F775W, F814W, F850LP
and WFC3 F098M, F105W, and F125W, is measured by running SExtractor in
dual-image mode with our WFC3 F160W mosaic as the detection image. All these
bands are smoothed by using the IRAF/PSFMATCH package\footnote[1]{We set the
PSFMATCH parameters ``filter''=``replace'' and ``threshold''=0.01 to replace
the very high-frequency and low signal-to-noise components of the PSF matching
function with a model computed from the low frequency and high signal-to-noise
components of the matching function.} to match their PSFs to that of the F160W
image. The WFC3 PSFs are generated by combining a core from TinyTim and wing
from stacked stars \citep[see][for details]{vanderwel12}, while the ACS PSFs
are generated by stacking stars. SExtractor is run on the PSF-matched image of
each band in both cold and hot modes with the same configurations as used for
the F160W photometry. Therefore, the source detection, segmentation area, and
isophotal area of sources are identical for the F160W and other \hst\ images.
The cold and hot photometry of each band is merged based on the F160W cold and
hot detection and combination.

Because the isophotal areas of a source in all \hst\ bands are identical, the
isophotal fluxes (FLUX\_ISO in SExtractor) can be used to measure colors among
the \hst\ bands. Under some circumstances, e.g., measuring stellar mass from
spectral energy distribution (SED) fitting, the total fluxes of all bands are needed. For these purposes, we
provide an inferred total flux for each band in the following way. For the
detection band (F160W), we use the photometry from within the Kron elliptical
aperture (FLUX\_AUTO in SExtractor) as the measure of total flux. We then
derive an aperture correction factor, $apcorr \equiv {\rm FLUX\_AUTO /
FLUX\_ISO}$, and apply it to 
other \hst\ bands to convert their isophotal fluxes and uncertainties into the
total fluxes and uncertainties: FLUX\_TOTAL = $apcorr \times$ FLUX\_ISO and
FLUXERR\_TOTAL = $apcorr \times$ FLUXERR\_ISO.
Our aperture correction method provides an accurate estimate of colors and
fluxes subject to the prior assumption that the PSF-convolved profile is the
same in all bands. While this assumption is not likely to hold for
well-resolved sources, most of the sources in the image are small enough that
any wavelength dependence of the galaxy profile will have very little impact on
the integrated flux ratio. The total flux and its uncertainty (FLUX\_TOTAL and
FLUXERR\_TOTAL) are provided in the catalog.

\begin{figure*}[htbp]
\center{
\hspace*{-2.3cm}
\includegraphics[scale=0.48, angle=0]{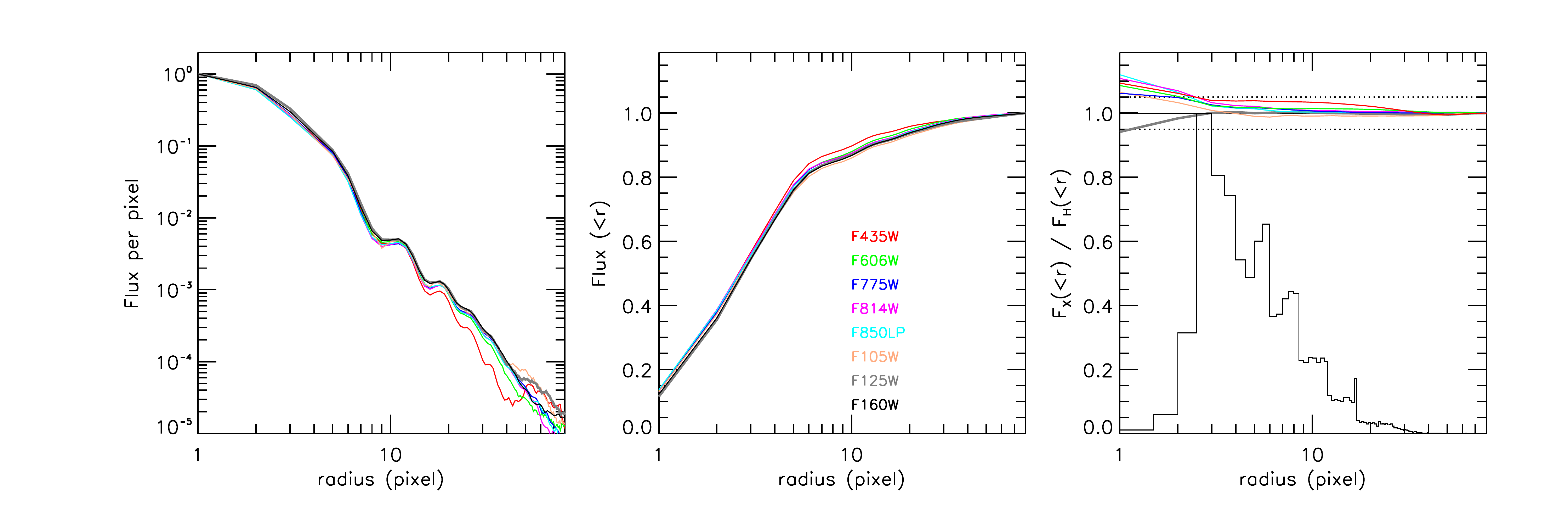}}
\caption[]{Accuracy of PSF matching between other \hst\ bands and F160W. {\it
Left}: the light profile of matched PSFs for each band. {\it Middle}: the curve
of growth of each matched PSF. {\it Right}: the curve of growth of each matched
PSF normalized by the curve of growth of the F160W PSF. In this panel, curves
with values greater than unity are under-smoothed, and vice versa. All
curves are color coded as labels in the middle panel show. Dotted lines in the
right panel show the 5\% relative error. The solid histogram in the right panel
shows the distribution of isophotal radii of all objects in our catalog.
\label{fig:psfcurve}}
\vspace{-0.2cm}
\end{figure*}

The accuracy of the PSF matching is crucial to our \hst\ photometry. To test
it, we extract stars from the CANDELS deep region (to ensure a relatively high
S/N) from all PSF-matched \hst\ images and compare their light profiles and
curves of growth in Figure \ref{fig:psfcurve}. 
The curves of growth of all \hst\  bands, normalized by that of the F160W band,
quickly converge to unity after a few pixels.
If the radius of the {\it isophotal aperture} of a source is larger than two
pixels (0\farcs12), the relative error of isophotal fluxes in all \hst\ bands
is less than 5\%. Overall, our PSF-matching does not induce a significant
systematic offset for the bulk of our sources.
Only 2.5\% of sources in our sample have isophotal radii less than 2 pixels.
For them, the relative photometric systematic offsets or uncertainties induced
by PSF-matching could be larger than 5\%. For these sources, we enforce a
minimum aperture with radius of 2.08 pixels (0\farcs125). Fluxes within the
minimum aperture (SExtractor parameter FLUX\_APER) instead of FLUX\_ISO are
used for these sources in all \hst\ bands. The aperture fluxes and
uncertainties are then scaled up by a new aperture correction factor, $apcorr
\equiv {\rm FLUX\_AUTO / FLUX\_APER}$ of the F160W band, to convert into the
total fluxes and uncertainties.

\subsection{HST Noise Properties}
\label{hstphoto:noise}

The uncertainties of our HST photometry are essentially the SExtractor
FLUXERR\_ISO parameter scaled by an aperture correction factor.  Recent works
\citep[e.g.,][]{wuyts08,coe13} have suggested that this parameter may
underestimate the true photometric uncertainties by a significant factor (up to
2--3), primarily due to the pixel-to-pixel correlations that are induced by the
step of drizzle \citep[e.g.,][]{casertano00} in the image reduction/combination
process.

However, in our case we provide SExtractor an RMS map for each filter that has
already been corrected for the correlations. Our pipeline that generates the
drizzled \hst\ images also produces a weight map whose value is nominally the
inverse variance of the expected background noise, predicted using a noise
model for the instrument that takes into account the background level, the
exposure time, and the expected instrumental noise. In principle, this weight
map should correctly describe the actual image noise. However, the noise in the
drizzled science images is suppressed by the pixel-to-pixel correlations. We
measure the background noise and the factor by which it is suppressed using an
IRAF script ``acall'', originally developed for GOODS (M. Dickinson, private
communication). The script runs on a relatively empty region of the image with
relatively uniform exposure time .
After masking our objects in the image and removing any low-level background
variations by median filtering on large angular scales, the script computes the
auto-correlation function of the unmasked pixels. 
The 2D auto-correlation image typically has a strong peak and falls off sharply
on a scale of a few drizzled pixels, since the data reduction (mainly
drizzling) does not introduce correlations on scales much larger than 1 or 2
original detector pixels. The script measures the peak value of the
auto-correlation image and the total power integrated out to some small radius
that can be set interactively. The square root of this ratio (peak over total)
is the RMS suppression factor ($C_s \leq 1$) due to the pixel-to-pixel
correlations. The measured inverse variance, corrected for the correlations, is
therefore $(C_s/{\rm RMS})^2$. This is compared to the value predicted from the
weight map produced by the drizzle pipeline, and if necessary the weight map is
re-scaled accordingly. The final RMS map used for photometry is then computed
from the inverse square root of the re-scaled weight map.

We carry out a test, equivalent to the ``empty aperture'' method of measuring
the variance within empty apertures of various sizes placed throughout the
images \citep[e.g.,][]{wuyts08,whitaker11,coe13}, to examine the noise
properties of our \hst\ images. First, we mask out all detected sources in the
\hst\ images. Then, for each band, we block sum both the drizzled science image
and the variance map (square of the RMS map) with various block sizes ($rb$, in
unit of pixel). Each block can be treated as an ``empty box''. We then
calculate the ratio of the background RMS of the summed science image and the
median of the summed RMS map (square root of the summed variance map). The
ratio should be equal to unity if the pixel-to-pixel correlations are
corrected.

\begin{figure}[htbp]
\center{
\hspace*{-0.7cm}
\includegraphics[scale=0.35, angle=0]{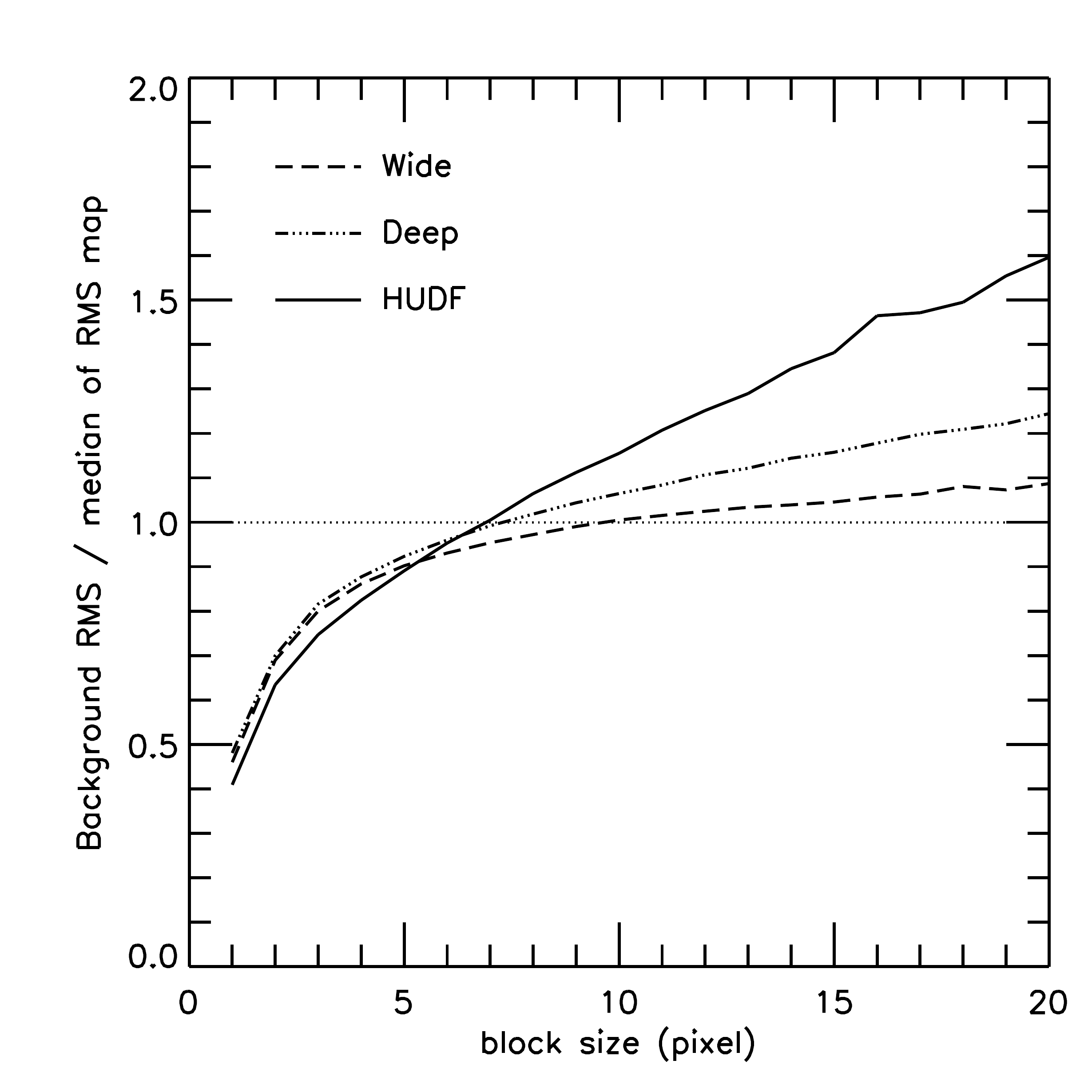}}
\caption[]{Ratio of the background RMS of block summed F160W image and the
median of the summed RMS map (square root of the block summed variance maps) as
a function of the block size.
\label{fig:rmscheck}}
\vspace{-0.2cm}
\end{figure}

Figure \ref{fig:rmscheck} shows the ratio as a function of $rb$ in the wide,
deep, and HUDF regions of the F160W image. The ratio is close to 1 at
5$<$rb$<$13, demonstrating that our RMS map represents the background noise
correctly on the scales relevant for most of our sources in the image, whose
typical sizes are $\sim$ 7 pixels.
At rb$>$13, the background RMS is larger than the median of the RMS map, mainly
due to the unmasked wings of detected objects and undetected sources, which
contaminate most blocked pixels when $rb$ is large.  It could also be due to
any larger scale fluctuations not removed by the flat fielding and sky
subtraction steps of the data processing.  This effect is strongest in the HUDF
region because it has the highest source number density and detects more
extended wings of objects than the other two regions.

The effect of correlation noise becomes more severe at smaller scale. At
$rb$=1, the background RMS is lower than the median of the RMS map by a factor
of 2$\sim$3, implying that the unblocked pixel-to-pixel RMS would significantly
underestimate photometric uncertainties due to the correlation. This result is
consistent with the test of \citet[][see its Figure 3(a)]{wuyts08}.  However,
our photometric uncertainties are not measured from a linear scaling of the
pixel-to-pixel RMS of the drizzled science image. Instead, they are measured
from the corrected RMS map, which is made from the re-scaled weight map to
represent the uncorrelated background fluctuation of our images. Therefore, we
conclude that our photometric uncertainties in F160W are not underestimated by
the pixel-to-pixel correlations. The same test on other HST bands shows similar
results. 

Last, it is important to note that the \hst\ photometric uncertainties in our
catalog are most appropriate when the fluxes are used to compute colors in
combination with other bands in this catalog. The \hst\ S/Ns in our catalog are
actually measured within the isophotal area and hence are higher than the total
S/Ns measured from a larger aperture (e.g., Kron radius). If the S/Ns in our
catalog are used instead as an estimate of the total S/Ns in these bands, the
photometric uncertainties should be inflated, roughly by the ratio of the F160W
S/N in the AUTO aperture to the S/N in the ISO aperture. 


\section{Photometry of Low-resolution Images}
\label{lowresphoto}

In a multi-wavelength survey, it has long been a challenge to measure reliable
and uniform photometry for bands whose spatial resolutions differ dramatically.
In GOODS-S, the FWHM of the PSFs varies from $\sim$0\farcs17 for \hst\ images to
$\sim$2\farcs0 for IRAC 8.0 $\mu$m, a factor of $\sim$12. Close pairs
identified in a high-resolution image could be blended into single sources in a
low-resolution image if the separation between the companions is less than 1
FWHM of the low-resolution image. In order to measure the flux of each member of
such pairs in the low-resolution image, a few methods have been used in
literature, e.g., the template fitting \citep{grazian06,laidler07}, the
``clean'' process migrated from radio astronomy \citep{wangweihao10}, etc. 

In our paper, we use a software package developed by the GOODS team, TFIT, to
carry out the template-fitting method to measure low-resolution photometry.
Details of TFIT are given by \citet{laidler07}, while thorough tests on its
robustness and uncertainties through multi-band simulations are given by
\citet{leeks12}.  For each object, TFIT uses the spatial position and
morphology of the object in a high-resolution image to construct a template.
This template is smoothed to match the resolution of the low-resolution image
and fit to the low-resolution image of the object. During the fitting, flux is
left as a free parameter. The best-fit flux is taken to be the flux of the
object in the low-resolution image, and the variance of the fitted flux is the
uncertainty in the flux. These procedures can be simultaneously done for
several objects which are close enough to each other in the sky so that the
blending effect of these objects on the flux measurement would be minimized.
Experiments on both simulated and real images show that TFIT is able to measure
accurate photometry of objects to the limiting sensitivity of the image. 

\begin{figure}[htbp]
\center{
\hspace*{-0.4cm}
\includegraphics[scale=0.33, angle=0]{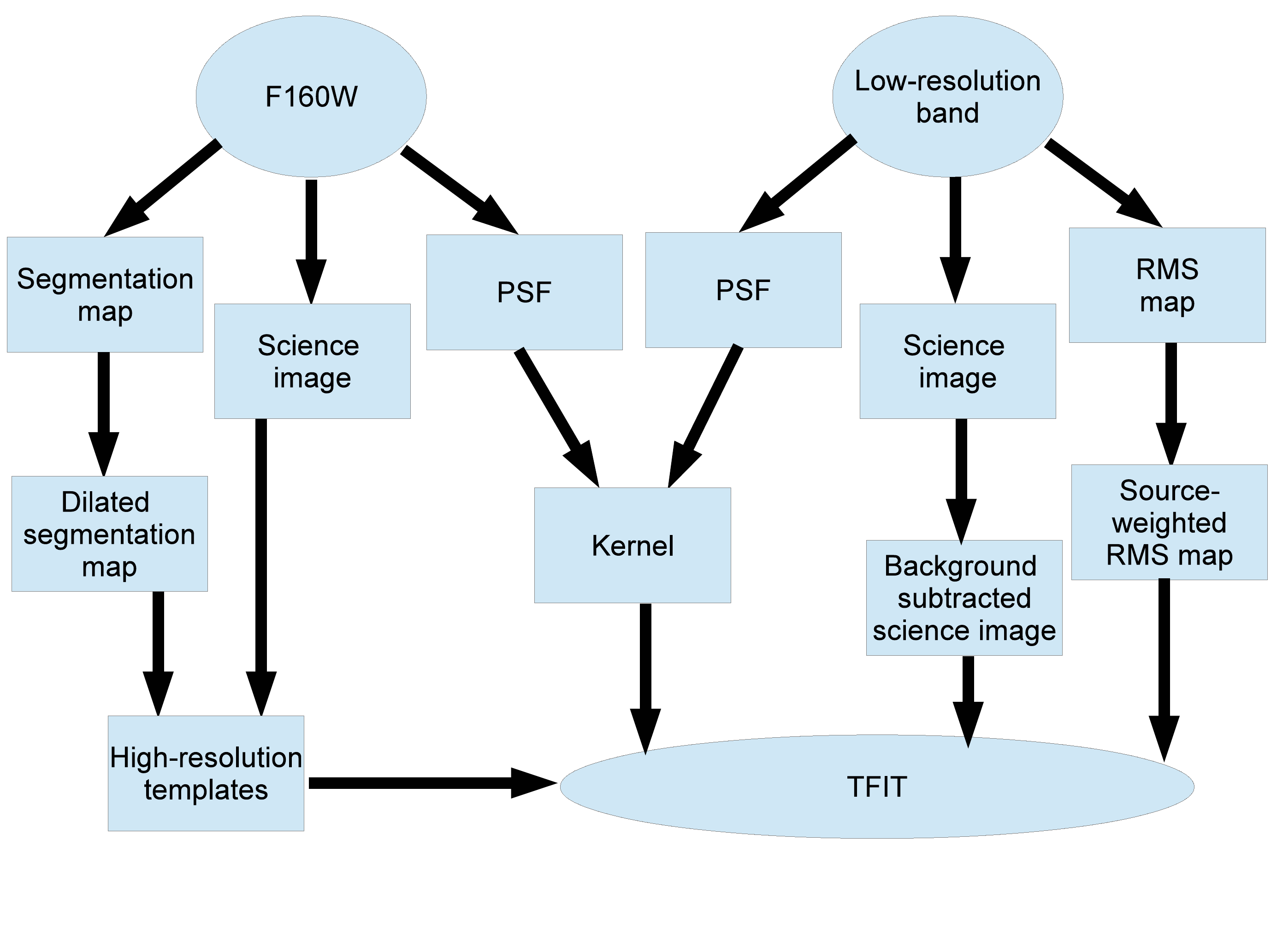}}
\caption[]{Flow chart of pre-processing of images prior to feeding them into TFIT for measuring photometry for low-resolution bands.
\label{fig:tfitflow}}
\vspace{-0.2cm}
\end{figure}

\begin{figure*}[htbp]
\center{
\hspace*{-0.0cm}
\includegraphics[scale=0.95, angle=0]{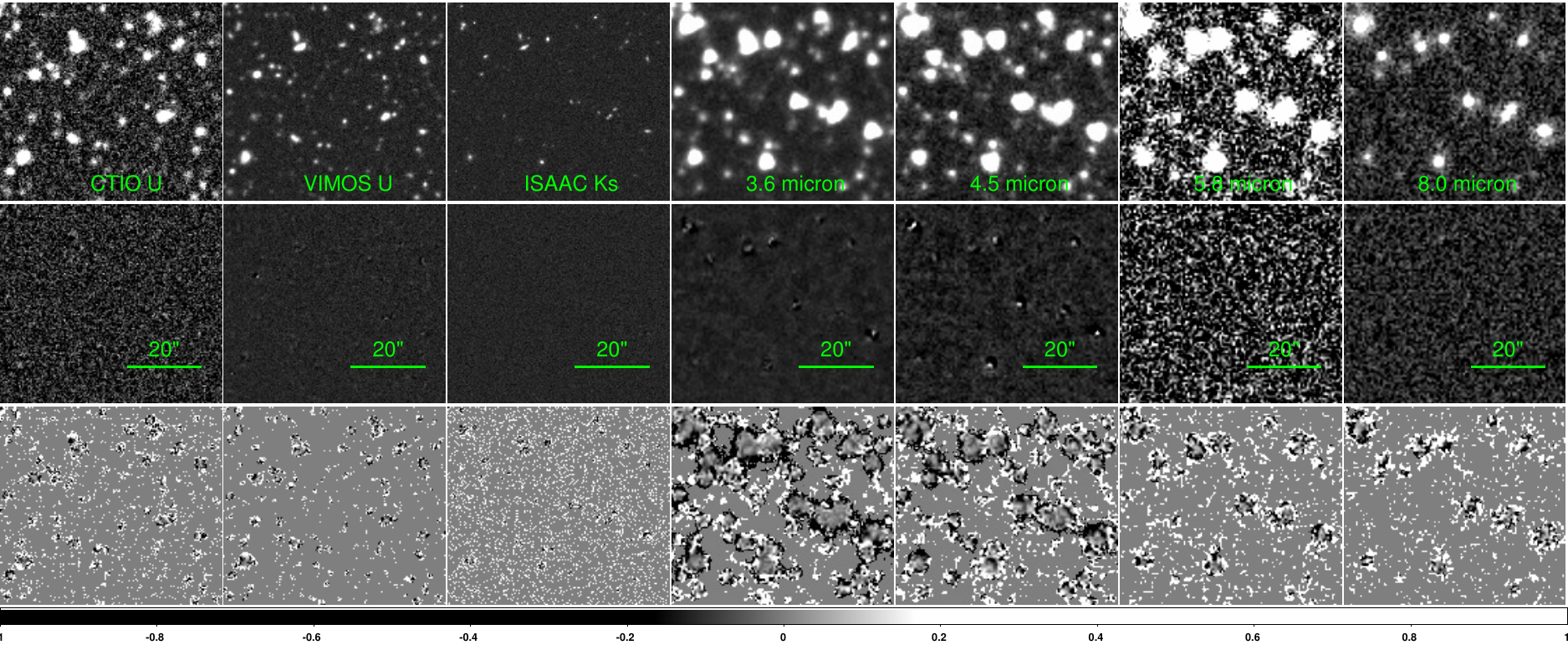}}
\caption[]{Example of the original image ({\it top}), the residual image after
TFIT procedure ({\it middle}), and the fraction of light left in the residual
image ({\it bottom}) of several low-resolution bands as indicated in the upper
panels in a representative sky region. The bottom panel is the ratio of the
middle and the top panels. The gray-scale bar shows the contrast of the bottom
panel. \\
\label{fig:resid}}
\vspace*{-0.5cm}
\end{figure*}

There are several steps for pre-processing both high-resolution and
low-resolution images prior to feeding them to TFIT. Figure \ref{fig:tfitflow}
presents a flow chart of these steps. \citet{galametz13uds} give a detailed
description of each step. Here we only summarize some key steps:

\begin{enumerate}

\item {\it High-resolution templates:} For each source detected in our F160W
mosaic, we cut out a postage stamp for it as the high-resolution template. The
size of the postage stamp is dependent on the segmentation area generated by
SExtractor. This segmentation area of a source, however, only contains pixels
above the isophotal detection threshold and could result in an artificial
truncation of the light profile of the object. \citet{galametz13uds} determined
an empirical relation to extend the segmentation area of an F160W source to a
proper size to include the outer wings of an object in the template. This
process, called ``dilation'', was thoroughly tested and optimized by
\citet{galametz13uds}.

\item {\it Convolution kernel:} TFIT requires a convolution kernel to smooth
the high-resolution templates to low-resolution. We use the IRAF/PSFMATCH
package to compute the kernel for ground-based bands (CTIO U, VIMOS U, and
ISAAC Ks). Our F160W PSF is a hybrid of the empirical PSF from stacking stars
and the simulated PSF from TinyTim \citep{vanderwel12}. Our ground-based
low-resolution PSFs and IRAC 3.6 and 4.5 $\mu$m PSFs are generated by stacking
stars, and IRAC 5.8 and 8.0 $\mu$m PSFs are model PSFs convolved with an
empirical kernel to slightly broaden them. For IRAC channels, with FWHMs more
than ten times broader than that of F160W, we simply use the IRAC PSFs as
kernels to smooth the F160W templates.

\item {\it Low-resolution images:} In principle, TFIT can fit both background
and objects simultaneously. However, we choose to subtract the background of
low-resolution images before running TFIT because fitting background would
introduce strong degeneracy and large uncertainty to the photometry of faint
sources. \citet{galametz13uds} give details of our background subtraction
algorithm. We also run the ``empty block'' test described in Sec.
\ref{hstphoto:noise} to examine the noise properties of the low-resolution
images. In order to correct the effect of the pixel-to-pixel correlations
induced by the image reduction, we re-scale the RMS map of each band so that
the ratio of the background RMS of the block summed image and the median of the
square root of the block summed variance map is equal to unity at block sizes
that are close to the typical sizes of sources in the image.

\item {\it Special treatment for individual bands:} Several low-resolution
bands need special treatment. For both the CTIO/MOSAIC and VLT/VIMOS U-bands,
if a source is covered by our ACS F435W image, we use the ACS F435W
instead of the WFC3 F160W image as the high-resolution template to minimize the
effect of morphological change from the UV to NIR wavelengths. To keep the same
detection sources, we still use the dilated F160W segmentation map for creating
high-resolution postage stamps. About 2300 sources ($\sim$7\% of our
catalog) are not covered by the ACS F435W image. For them, the F160W image is
used as the template for both U-bands. For VLT/ISAAC Ks-band, we run TFIT on
each of its tiles instead of on the co-added mosaic because the tile-by-tile
variation of PSF is non-negligible. We also run TFIT for each epoch of IRAC 5.8
and 8.0 $\mu$m due to the PSF variation. For sources that are fitted more than
once in the ISAAC and IRAC bands, we calculate a weight-averaged flux from each
fitting using the squared fitting uncertainty as the inverse weight.

\end{enumerate}

To correct the geometric distortion and/or mis-registration between the high-
and low-resolution images, we run TFIT in two passes. In the first pass, TFIT
calculates a cross-correlation between the model image and the low-resolution
image to determine the optimal shifted transfer kernels for each zone in the
image. The second pass then uses these shifted kernels to smooth the
high-resolution templates to the low resolution. The two-pass scheme
effectively reduces the mis-alignment between templates and low-resolution
images.

An example of the original images and residual maps of some TFITed bands is
shown in Figure \ref{fig:resid}. A simple visual comparison between the
original and the residual images suggests that TFIT does an excellent job of
extracting light for the majority of sources in the low-resolution bands.  The
overall residual is quite close to zero in the maps of U, Ks, and IRAC 5.8 and
8.0 $\mu$m. Some (especially bright) objects have obvious residuals left in the
map of IRAC 3.6 and 4.5 $\mu$m. To quantify the residual, we calculate the
fraction of light left in the residual maps (bottom panels). For most pixels in
the sources with the worst residuals, the residual light is around $\pm$10\% of
the original signal. The average residual over the area of these sources,
taking into account both positive and negative pixels, is actually quite close
to zero, suggesting that the photometry of bright sources in IRAC 3.6 and 4.5
$\mu$m is not significantly mis-measured, even though their residual maps look
worse than others. The residual maps only provide a rough estimate of the
quality of our photometry. A quantitative analysis on the quality of the
catalog will be presented in Sec.  \ref{quality}.

After measuring the low-resolution photometry with TFIT, we merge the
photometry of all the available bands, both low-resolution and \hst, into the
final catalog. This step is straightforward because each source keeps its ID
and coordinates as determined in the F160W detection through the whole
procedure of measuring its multi-band photometry. For each source, we also
include a weight in each band. For \hst\ and \spitzer\ bands, the weight is the
mean exposure time within six pixels from the center of the object. For other
bands, it is the average weight within six pixels from the center. The column
descriptions of the catalog are given in Table \ref{tb:column}.

\section{Quality Checks}
\label{quality}

We test the quality of our catalog by: (1) comparing colors of stars in our
catalog and that in stellar libraries, (2) comparing our catalog with other
published catalogs in GOODS-S, (3) measuring zeropoint offsets through the
best-fit templates of the SED of
spectroscopically observed galaxies, and (4) evaluating the accuracy of
photometric redshifts (photo-zs) measured with our catalog.

\subsection{Colors of Stars}
\label{quality:star}

\begin{figure}[htbp]
\center{
\hspace*{-1.0cm}
\includegraphics[scale=0.22, angle=0]{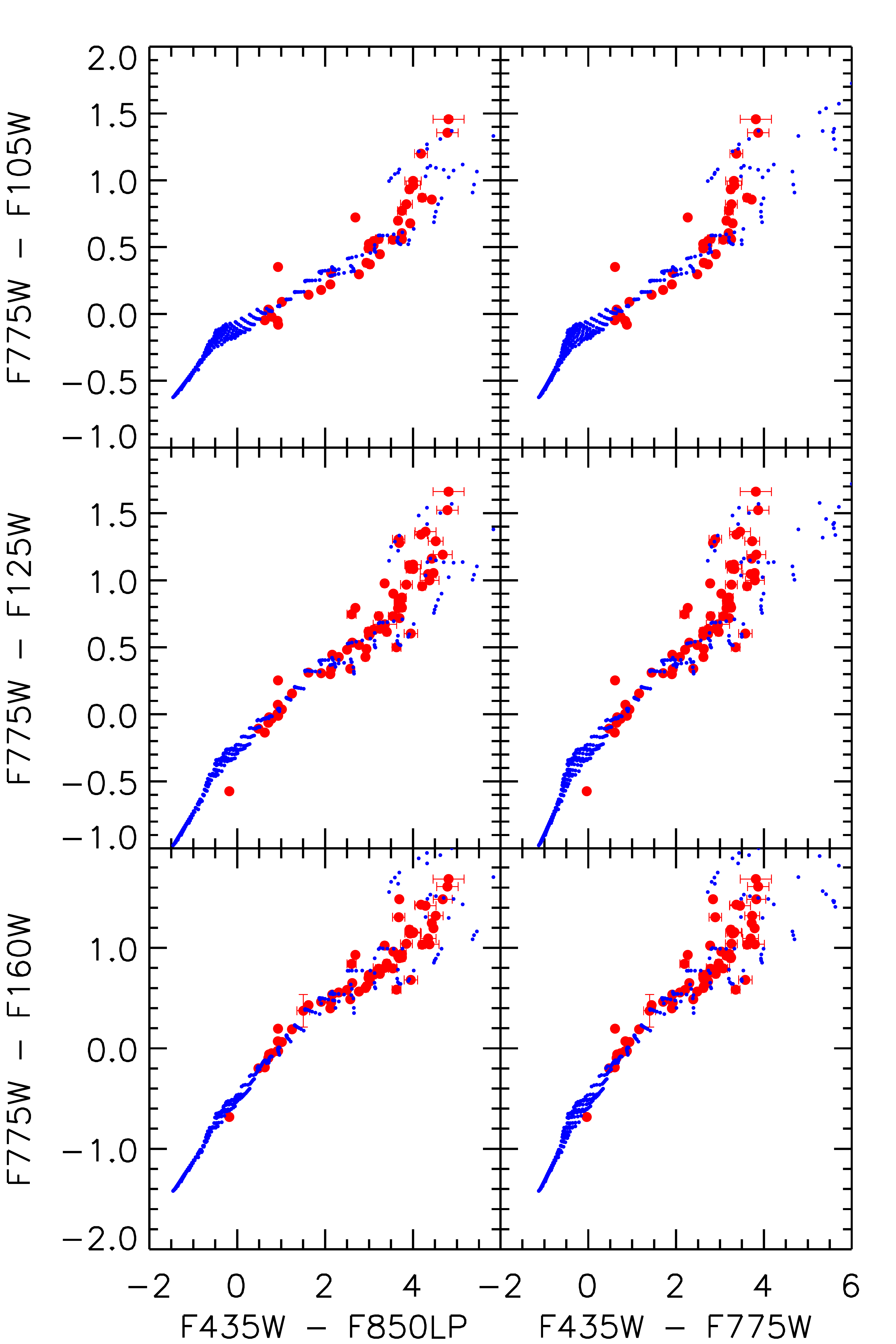}}
\vspace*{0.05cm}
\caption[]{Comparison between colors of stars in our catalog (red points with
error bars) and that of stars in the synthetic library of \citet{lejeune97} (blue). Colors are in AB magnitudes as indicated.
\label{fig:hstcolor}}
\vspace*{-0.2cm}
\end{figure}

We first check the quality of our \hst\ photometry by comparing the ACS--WFC3
colors of stars in our catalog with that of stars in stellar libraries.
Because GOODS-S is located at high Galactic latitude, a large fraction of its
stars may be metal-poor halo stars. The ACS--WFC3 colors, namely, the
optical--NIR colors, of stars strongly depend on the metallicity of the stars
because the optical fluxes are heavily absorbed by metals while the NIR fluxes
are much less affected.  Therefore, the ACS--WFC3 colors of stars in GOODS-S
may be systematically different from the colors of stars in some commonly used
stellar libraries, which are mainly calibrated with bright disk stars with
richer metallicity. We use the model of stellar population synthesis of our
Milky Way of \citet{robin03}\footnote{http://model.obs-besancon.fr} to estimate
the metallicity distribution of stars in GOODS-S and find a median value of
[Fe/H]$\sim$-0.5. We then choose the set of models with metallicity [M/H]=-0.5
from the synthesis stellar library of \citet{lejeune97} as our standard
reference of stellar colors.

Figure \ref{fig:hstcolor} shows the locations of stars in our catalog and those
in the Lejuene library in ACS--WFC3 color-color diagrams.  Stars are selected
with the SExtractor parameter CLASS\_STAR$>$0.98 in the F160W band. For each
color, we only use stars with S/N$>$3 in the two bands used to calculate the
color.
If the stellar photometry of one WFC3 band (denoted by X)  is incorrectly
measured, the (F775W--X) color of observed stars would deviate vertically 
from that of library stars in {\it both} (F775W--X) vs. (F435W--F850LP) {\it
and} (F775W--X) vs.  (F435W--F775W) diagrams by a similar amount. In another
situation, if the photometry of one of the ACS Biz bands is mis-measured, the
observed stellar color would deviate horizontally by a similar amount from
that of library stars in {\it all} diagrams of (F775W--X) vs. ACS color, where
the ACS color involves the problematic band. There is excellent agreement
between our observed stars and library stars for all colors in the plot,
suggesting that our stellar ACS--WFC3 colors are accurate.

\subsection{Comparison with Other Catalogs}
\label{quality:others}

\begin{figure*}[htbp]
\center{
\hspace*{-0.5cm}
\vspace*{-0.5cm}
\includegraphics[scale=0.40, angle=0]{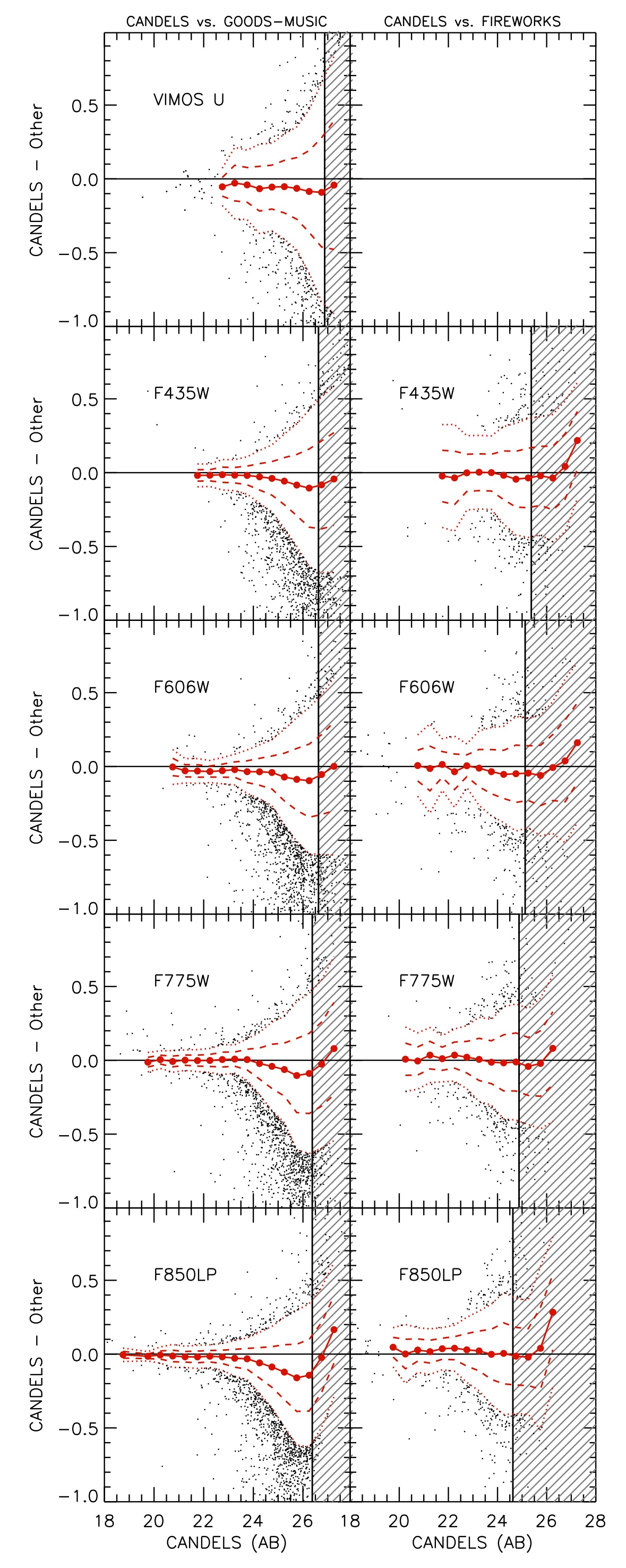}
\includegraphics[scale=0.40, angle=0]{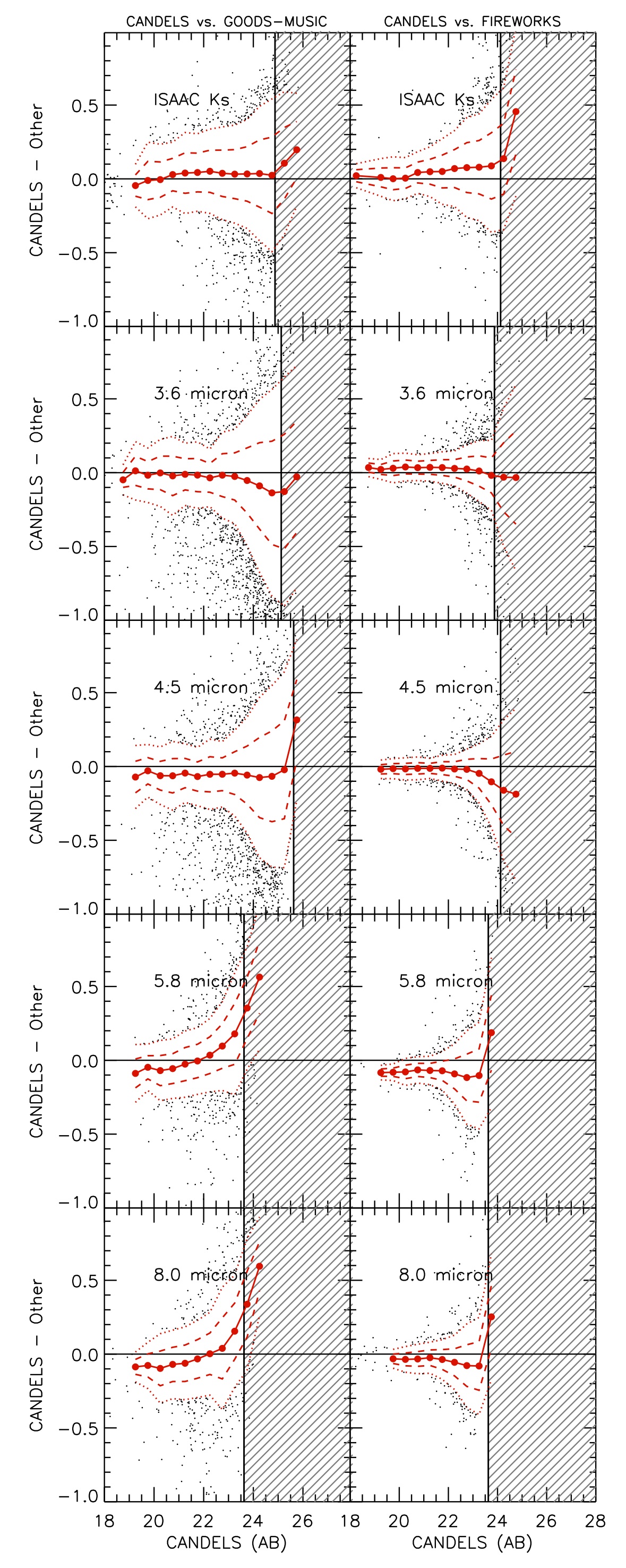}
\vspace*{0.5cm}
\caption[]{Comparison between our photometry (``CANDELS'') and two published
catalogs (``Other'') in GOODS-S: GOODS-MUSIC (first and third columns) and
FIREWORKS (second and fourth columns). For each band, we only use sources with
S/N$>$3 in both our and the other catalog for comparison. In each panel, red
points connected by solid line show the mean (after 3$\sigma$-clipping) of the
magnitude difference (defined as CANDELS - Other) as a function of magnitude.
Upper and lower red dash/dotted lines show 1$\sigma$/2$\sigma$ confidence level
after 3$\sigma$-clipping of the magnitude difference. Black dots show objects
whose magnitude difference is beyond 2$\sigma$ of the mean. The shaded area of
each panel is subjected to the Eddington bias (see the text and discussion on
Figure \ref{fig:tfitvsseds} for details) so the comparison there is biased. \\
\label{fig:crosscat}} 
}
\end{figure*}


We also compare our photometry with other published multi-wavelength catalogs
in GOODS-S. The comparison helps in two ways. First, it provides an assessment of
the photometry of extended sources. Second, although we find no systematic
offset in our stellar ACS--WFC3 colors, we still need an absolute measurement
of the fluxes of either ACS or WFC3 bands.

Two multi-wavelength catalogs in the literature cover (almost) the entire
GOODS-S field as well as include most of our bands: GOODS-MUSIC
\citep[GM,][]{grazian06,santini09} and FIREWORKS \citep[FW,][]{wuyts08}. GM
includes ACS F435W, F606W, F775W, and F850LP images, VLT JHKs data,
\spitzer/IRAC (3.6, 4.5, 5.8 and 8.0 $\mu$m) and MIPS (24 $\mu$m) data, and
publicly available U-band data from the ESO 2.2-meter telescope and VLT/VIMOS.
A software package, ConvPhot, which operates on the same principle as TFIT, but
is a completely independent implementation, has been developed for measuring
PSF-matching photometry for space and ground-based images of different
resolutions and depths. Different from our detection scheme, GM detects objects
mainly in F850LP and is supplemented by Ks- and 4.5 $\mu$m-band detected
sources.

FW is a Ks-selected catalog for the CDF-S,
containing photometry in the ACS F435W, F606W, F775W, and F850LP, ground-based
U, B, V, R, and I, VLT J, H, and Ks, IRAC 3.6, 4.5, 5.8, and 8.0 $\mu$m, and
MIPS 24 $\mu$m bands. Photometry in ACS and ISAAC bands is measured using
SExtractor in dual-image mode with the Ks-band mosaic as the detection image.
Color and aperture photometry are measured with the same apertures in each
band. SExtractor MAG\_AUTO is used to derive the total flux of the Ks-detected
objects. An aperture correction is applied to compute the total integrated flux
based on the curve of growth of the PSF of the Ks-band. The total-to-aperture
correction factor of the Ks band is then applied to the other ACS and ISAAC
bands.  Photometry in IRAC bands is measured with the method of
\citet{labbe06}.  Similar to TFIT, it fits the light profile of the higher
resolution image to that of lower resolution images and leaves flux as a free
parameter. For a single source, however, the method does not use the best-fit flux as
the total flux of the source. Instead, it subtracts its neighbors based on the
best-fit models and then estimates the flux of the source of interest via
aperture photometry.

\begin{figure*}[ht]
\center{
\hspace*{-0.5cm}
\vspace*{-0.5cm}
\includegraphics[scale=0.40, angle=0]{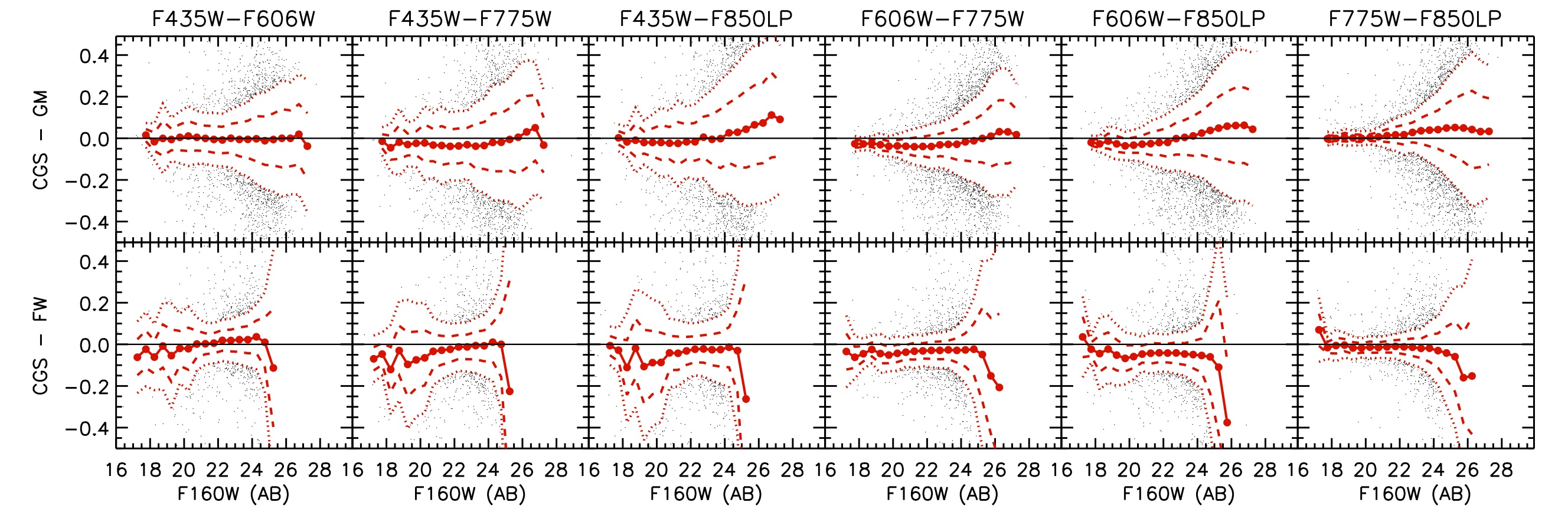}
\vspace*{0.5cm}
\caption[]{Differences of ACS colors measured by our catalog (CGS) and
GOODS-MUSIC (GM, upper panels) and FIREWORKS (FW, lower panels). In each panel,
red points connected by solid line show the mean (after 3$\sigma$-clipping)
difference of one ACS color in CGS and the other catalog as a function of the
F160W magnitude. Upper and lower red dash/dotted lines show 1$\sigma$/2$\sigma$
confidence level after 3$\sigma$-clipping of the magnitude difference. Black
dots show objects whose magnitude difference is beyond 2$\sigma$ of the mean.
We only use sources with S/N$<$5 in ACS and F160W bands in the comparison.
\label{fig:crosscolor}}
}
\end{figure*}

In order to compare our photometry with that in the two catalogs, we match our
CANDELS GOODS-S catalog (CGS) to the other two through source coordinates with a maximum
matching radius of 0\farcs3 and 0\farcs5 for GM and FW, respectively. We only
use cleanly detected sources (i.e., not saturated, not truncated, not having
bad pixels, etc.) in our catalog with the F160W SExtractor parameter FLAG=0.
The comparisons are shown in Figure \ref{fig:crosscat}. For each band, we only
consider objects with S/N$>$3 in our and the other catalogs. 

On average, there is a good agreement with nearly zero systematic offset
between CGS and the other two catalogs over the magnitude range to $\sim$ 24 AB
mag in most bands (except the IRAC 5.8 and 8.0 $\mu$m). The offset reaches
$\sim$ 0.05 mag in the worst bands. At the faint end ($>$ 24 AB mag),
deviations are evident between CGS and the other two. 

At the very faint end (the shaded areas in Figure \ref{fig:crosscat}), the
Eddington bias makes the comparison unreliable. The flux uncertainties of
objects in a shallower catalog (GM and FW) are larger so that objects in it are
more likely to be scattered to brighter or fainter magnitudes than in a deeper
catalog. Sources close to the S/N=3 limit in CGS should be included in the
comparison. In fact, however, if the fluxes of their counterparts in GM or FW
are scattered toward fainter fluxes, 
these sources are excluded because their S/N in GM or FW is now $<$3.
Therefore, at the S/N limit, only those CGS sources whose GM and FW fluxes are
scattered toward brighter fluxes 
are included in the comparison. This biases the comparison. In order to
determine the magnitude range subjected to Eddington bias in each band, we fit
a power law to the differential number count density of GM and FW without any
S/N cut. After the S/N$>$3 cut being applied, magnitude ranges where the new
differential number count density is less than 50\% of the best-fit power law
are now very incomplete in GM and FW and hence induce Eddington bias. The
comparisons in these magnitude ranges are not discussed here because they are
unreliable.

Finally, all comparisons in this section only show the
consistency/inconsistency between our and the other two catalogs. They cannot
tell us whether one catalog is more accurate than the others.

\subsubsection{UV to Optical Photometry}
\label{quality:others:uvoptical}

There is an almost constant offset of $\sim$-0.05 mag between CGS and GM in the
VIMOS U-band to $\sim$27 AB mag, where CGS measures brighter magnitudes. This
deviation could be caused by the different high-resolution templates and
kernels used in the template profile fitting method of both catalogs. In GM,
ACS F850LP images are used as the templates to fit the U-band image, while in
CGS, ACS F435W images are used to minimize the morphological changes along with
wavelength. To provide an independent check, we compare our CGS photometry with
the SExtractor MAG\_AUTO photometry of \citet{nonino09}. We find good
agreement with almost zero offset over the magnitude range to $\sim$27 AB mag. 

In the ACS bands, the agreement between our and the other two catalogs
is quite good. The offset between CGS and GM is almost zero for objects with
magnitude brighter than 24 AB mag. The largest offset, seen in the ACS F606W
band, is about -0.02 mag. For objects fainter than 24 AB mag, CGS magnitudes
are brighter, and the deviation between the two catalogs increases toward
fainter magnitude.
The deviation may be due to different apertures used to measure the total
fluxes in CGS and GM. The aperture size, represented by the KRON\_RADIUS in
SExtractor, is
defined in the F160W band in CGS and then applied to other \hst\ bands through
our aperture correction factors, assuming no morphological changes between
these bands. In GM, however, the KRON\_RADIUS is defined in the ACS F850LP
band. Due to the high sensitivity and broader PSF of WFC3, the KRON\_RADIUS defined
in the F160W band is typically larger than that defined in the F850LP band for
each individual source. Therefore, the F160W KRON\_RADIUS counts more light
from the wings of each object and measures a brighter magnitude. In fact, we
find an obvious correlation between the difference of magnitudes and the
difference of KRON\_RADIUS. Objects whose F160W KRON\_RADIUS is 
larger than the F850LP one are also brighter in our catalog, and vice versa.
This correlation supports our speculation that KRON\_RADIUS defined in
different images is the reason for the deviation in the faint end between CGS
and GM. More supportive evidence is from the excellent agreement between the
ACS colors of CGS and GM (the upper panels of Figure \ref{fig:crosscolor}).
Colors from both catalogs have almost zero ($\sim$0.02 mag in the worst case for
sources brighter than 26 AB mag in F160W) offset over the magnitude range to
F160W $\sim$ 26 AB mag. This excellent agreement demonstrates that our PSF
matching process does not induce a systematic offset to our ACS photometry.
Therefore, the offset in the faint end of ACS photometry between the two
catalogs is likely caused by the different apertures used to measure fluxes.

The comparison between CGS and FW shows large scatter, possibly due to both
small number statistics and different photometry measurements. FW has a faint
limit cut on 24.3 AB mag in its detection Ks band, which excludes the majority
of sources fainter than 25 AB mag in the ACS bands in our catalog. Therefore,
the comparison fainter than 25 AB mag has large uncertainty due to small number
statistics. Also, FW was generated based on an early version of the GOODS-S
ISAAC Ks-band image, which covers $\sim$130 arcmin$^2$, only 75\% of the
whole GOODS-S region. The smaller sky coverage further reduces the number of
matched sources in this comparison. For sources brighter than 26 AB, the
comparison has larger scatter, $\sim$0.15--0.2, than that in the CGS vs. GM
comparison. The large scatter is likely due to the treatment of isolated
sources and blended sources in the Ks-band in FW. For isolated sources, FW used
MAG\_AUTO as the total flux with KRON\_RADIUS defined in Ks-band. Because the
KRON\_RADIUS defined in the Ks-band is typically larger than that defined in
the \hst\ bands, isolated sources tend to have brighter magnitudes in FW than
in CGS. On the other side, to reduce the influence of neighboring sources, FW
used a reduction factor to shrink the isophotal area of blended sources and
used the isophotal flux to derive the total flux. Therefore, the fluxes of
blended sources are likely underestimated in FW. The combination of the above
two introduces large scatter to our comparison, as the plot shows. But overall,
the mean difference between FW and CGS is close to zero over the magnitude
range to $\sim$25 AB mag, where the Eddington bias begins to affect the
comparison.


\subsubsection{IR Photometry}
\label{quality:others:ir}

In the ISAAC Ks-band, CGS shows a magnitude-dependent deviation from both GM
and FW. CGS becomes fainter than the other two at $\sim$20 AB mag.  The
deviation increases as the flux of sources decreases. At $\sim$24 AB mag, the
deviation reaches to $\sim$0.05 mag in CGS vs. GM and $\sim$0.1 mag in CGS vs.
FW.
Due to the lack of sources with S/N$>$5 at magnitude fainter than 24 AB mag,
the deviation in the very faint end cannot be investigated. The deviation could
be due to over-subtraction of background in the ISAAC images during our
pre-processing of images (Sec. \ref{lowresphoto}). However, measuring global
background from the images used by TFIT shows no significant negative
background, suggesting that our background subtraction pipeline does a fairly
good job. The deviation in CGS vs. FW could also be due to the under-deblending
of neighboring sources in SExtractor-type photometry. Contaminating light from
poorly separated companions would boost the photometry of faint sources but
have little effect on bright sources. 

CGS shows good agreement with GM and FW over the magnitude range to 24 AB mag
in IRAC 3.6 and 4.5 $\mu$m. The deviation between CGS and GM is almost zero at
3.6 $\mu$m and about -0.04 mag at 4.5 $\mu$m, while that between CGS and FW is
about 0.03 mag at 3.6 $\mu$m and almost zero at 4.5 $\mu$m. The absolute
uncertainty in the IRAC calibration is about 3\%. 
Therefore, we believe that within the calibration uncertainty, there is no
major concern on our photometry in the magnitude range of 18--24 AB mag for the
IRAC 3.6 and 4.5 $\mu$m.


The comparison in the IRAC 5.8 and 8.0 $\mu$m can only be done for sources
brighter than 23 AB mag. The deviations between CGS and GM in both channels
are quite similar and  magnitude dependent. CGS is brighter than GM by
$\sim$0.1 mag at 19 AB mag and becomes fainter than GM by $\sim$0.05 mag at 23
AB mag. In contrast, the deviations between CGS and FW are almost constant,
with CGS being brighter by $\sim$0.07 mag and $\sim$0.03 mag in the 5.8 and 8.0
$\mu$m bands, respectively.


Although all three catalogs use the profile template fitting method to derive
the photometry of IRAC bands, the implementation of this method in each catalog
is quite different. The scatter in CGS vs. FW is significantly smaller than
that in CGS vs. GM. We suspect that the difference in scatter originates from
the choice of high-resolution templates. NIR images are used in both CGS
(F160W) and FW (Ks-band) as the templates for fitting IRAC images. Although the
resolution differs between the two sets of templates, the morphological change
from them to IRAC bands would be quite small. On the other side, GM uses the
ACS z-band as templates to fit the profile of sources in IRAC bands. Since the
templates and the IRAC images sample two distinct wavelength regimes, the
morphological change could induce large difference in photometry, including
systematic offset in the 5.8 and 8.0 $\mu$m bands, where the wavelength
difference between templates and sources becomes the largest.

\begin{figure}[htbp]
\center{
\includegraphics[scale=0.60, angle=0]{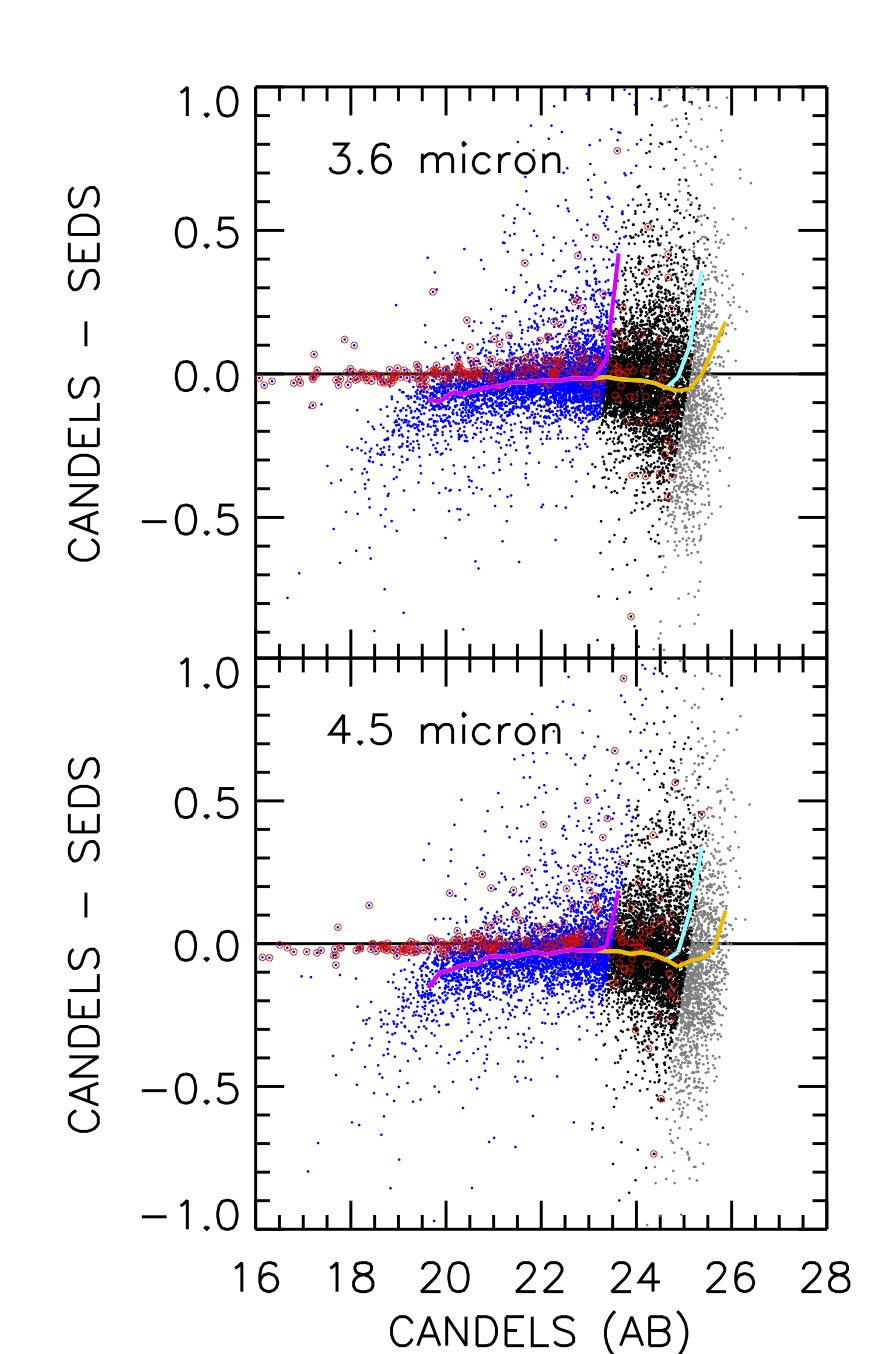}
\caption[]{Comparison between CANDELS TFIT photometry and SEDS photometry for
IRAC 3.6 ({\it upper}) and 4.5 $\mu$m ({\it lower}) bands. Points with different
colors represent sources with various S/N cuts on both catalogs: S/N$>$8
(blue), S/N$>$5 (black), and S/N$>$3 (gray). Red circles show point-like
objects with S/N$>$5 and F160W SExtractor parameter CLASS\_STAR$>$0.8. The
average differences of the comparisons with various S/N cuts are shown by
colored curves: violet for S/N$>$8, cyan for S/N$>$5, and yellow for S/N$>$3.
The large deviation at the faint end of each curve demonstrates the Eddington
bias caused by the imposed S/N cuts on both catalogs.
\label{fig:tfitvsseds}}
}
\end{figure}

We also compare our TFIT photometry to the official SEDS catalog
\citep{ashby13seds} for IRAC 3.6 and 4.5 $\mu$m bands. Ashby et al. used
StarFinder to fit PSFs to sources and subtracted the best-fit PSFs from the
original image. Then, for each source, they added its best-fit PSF back to the
residual map while keeping other sources subtracted. They measured aperture
photometry for the source with a variety of aperture sizes (here we choose the
one with size of 2\farcs4) and applied a correction factor to convert the
aperture photometry to the total photometry based on the curve of growth of the
IRAC PSFs. The comparison of CANDELS TFIT photometry and SEDS photometry is
shown in Figure \ref{fig:tfitvsseds}. For point-like sources (F160W SExtractor
parameter CLASS\_STAR$>$0.8) in both IRAC channels, the agreement between both
catalogs is excellent over the magnitude range to 24.5 AB mag with an offset of
$\sim$-0.04 mag. For bright ($<$20 AB mag) extended sources, however, SEDS
underestimates their fluxes. This is because the aperture correction used in
SEDS is only valid for point sources. When using larger apertures (e.g., size
of 6\farcs0), SEDS photometry of bright extended sources agrees with our TFIT
photometry much better with deviation $\sim$0.05 mag. \citet{ashby13seds} give
a detailed description and analysis of SEDS photometry and its uncertainty.

Figure 12 also illustrates the Eddington bias in comparing two catalogs.
Sources near the threshold will randomly have greater S/N in one catalog than
the other. Imposing a S/N cutoff means that sources just above the cutoff will
on average have catalog fluxes that are larger than the true values. If the
uncertainties of the two catalogs differ, the amount of bias will differ. This
effect can be seen in the figure as the sloping transition from blue (S/N$>$8)
to black (S/N$>$5) and from black (S/N$>$5) to gray (S/N$>$3) points. The
deviation of $\sim$0.05 mag at 23.5 AB mag for curves with S/N$>$8 is strongly
biased and hence does not suggest any true difference between the two
catalogs at this magnitude. In fact, good agreement between the two
catalogs can be seen down to 25.5 AB mag where the S/N$>$3 cut induces the
Eddington bias.

Due to the Eddington bias, it is difficult to evaluate the accuracy of our IRAC
photometry at the very faint end through comparisons with other catalogs.
Instead, we use fake source simulations to test our photometry at the detection
limit of the IRAC channels. We generate 1000 fake F160W sources and place them
randomly into our F160W mosaic. We then use our TFIT pipeline to measure the
IRAC photometry of these fake sources and real sources simultaneously. Because
there are no counterparts of these fake sources in IRAC images, on average
their IRAC photometry should be zero. Indeed, the distributions of the ratio of
IRAC photometry and its uncertainties of these fake sources in all IRAC
channels have a mean of zero and a standard deviation of one. This result
demonstrates that -- assuming no PSF or registration mismatches -- our catalog
accurately measures IRAC photometry and its uncertainties down to the detection
limits of IRAC images.

\subsection{Zeropoint Offsets}
\label{quality:zp}

To further assess the overall accuracy of our photometry and its influence on
deriving properties of galaxies, we measure the zeropoint offset of each band
by comparing the SEDs of spectroscopically observed galaxies to synthetic
stellar population models. If a band suffers from significant systematic bias,
its photometry will statistically deviate from the best-fit models.

The spectroscopic sample used in our study is from \citet{dahlen10}.  We match
the sample to our F160W band coordinates with a maximum matching radius of
0\farcs3. We only use galaxies with good \citep[flag=1 in ][]{dahlen10}
spectroscopic redshifts (spec-zs) and exclude X-ray detected sources. The
sample contains 1338 galaxies. The synthetic stellar population models used in
this test are retrieved from the PEGASE v2.0 library \citep{pegase}. Details of
SED-fitting can be found in \citet{ycguo12vjl}. For each galaxy, we fix its
redshift to its spec-z during the fitting.

\begin{table}
\begin{center}
\caption{Zeropoint Offset Derived from Best-fit SED Models \label{tb:zpoff}}
\begin{tabular}{ccc}
\hline\hline
Band  &   Zeropoint Offset\footnote[1]{Offsets are defined as observed magnitude minus model magnitude. A positive offset means our photometry is fainter than models, and vice versa.} 
  & Zeropoint Offset \\
      &   (Before Template Correction)  & (After Template Correction) \\
\hline
U\_CTIO & $0.03\pm0.07$ & $0.03\pm0.07$ \\
U\_VIMOS & $-0.01\pm0.06$ & $-0.01\pm0.06$ \\
F435W & $-0.03\pm0.08$ & $-0.03\pm0.08$ \\
F606W & $-0.01\pm0.05$ & $-0.01\pm0.05$ \\
F775W & $-0.01\pm0.05$ & $-0.01\pm0.04$ \\
F814W & $-0.01\pm0.05$ & $-0.01\pm0.05$ \\
F850LP & $-0.03\pm0.05$ & $-0.02\pm0.04$ \\
F098M & $0.02\pm0.05$ & $0.03\pm0.04$ \\
F105W & $0.01\pm0.04$ & $0.01\pm0.04$ \\
F125W & $0.01\pm0.05$ & $0.01\pm0.03$ \\
F160W & $0.02\pm0.05$ & $0.00\pm0.04$ \\
ISAAC Ks & $0.08\pm0.12$ & $0.06\pm0.12$ \\
HAWK-I Ks & $0.02\pm0.08$ & $0.01\pm0.07$ \\
IRAC 3.6$\mu$m & $-0.09\pm0.10$ & $-0.07\pm0.10$ \\
IRAC 4.5$\mu$m & $-0.08\pm0.12$ & $-0.03\pm0.10$ \\
IRAC 5.8$\mu$m & $-0.11\pm0.23$ & $-0.04\pm0.21$ \\
IRAC 8.0$\mu$m & $-0.31\pm0.49$ & $-0.20\pm0.47$ \\
\hline
\vspace*{-0.5cm}
\end{tabular}
\end{center}
\end{table}

The template fitting can suffer from systematic errors if the models do not
match real galaxy SEDs or if the absolute calibrations of the observations are
systematically wrong at different wavelengths. In order to correct for such
errors, we shift the observed and best-fit SEDs of each galaxy to rest-frame
wavelength and calculate the offset between them as a function of rest-frame
wavelength. The average offset of the spec-z sample at a given rest-frame
wavelength is thus contributed by multiple bands that observe galaxies at
different redshifts. Therefore, any offset found in such a way is likely due to
the inaccuracy of the SED templates instead of the systematic errors of our
photometry, unless all the bands contributing to the rest-frame wavelength
underestimate/overestimate fluxes in the same direction. We then subtract the
offsets at all rest-frame wavelengths from our SED templates. 

The zeropoint offsets of all bands are shown in Table \ref{tb:zpoff}, both
before and after applying the template correction.  
For most non-IRAC bands, the offset is $\sim$0.02 mag. The only non-IRAC band
with significant offset is ISAAC Ks-band, whose offset is 0.06 mag. For IRAC
3.6, 4.5, and 5.8 $\mu$m, the offsets are reduced by the template correction to
$\sim$0.05 mag. This value is consistent with the deviation between different
measurements from our previous comparisons with other catalogs. The offset of
IRAC 8.0 $\mu$m is quite large, though. For galaxies at z$<$0.4 (about 20\% of
our spec-z galaxies), this band is affected by the emissions of polycyclic
aromatic hydrocarbon (PAH). The PAH emissions are not characterized in our SED
models even after our correction, and thus the observed IRAC 8.0 $\mu$m flux
should be much brighter than the SED templates.

\begin{table*}[htbp]
\begin{center}
\caption{Accuracy of Photometric Redshift \label{tb:photoz}}
\vspace*{-0.3cm}
\begin{tabular}{r||cccc||cccc||cccc}
\hline
\hline
\multicolumn{1}{r||}{} & \multicolumn{4}{c||}{All} & \multicolumn{4}{c||}{z$\leq$1.5} & \multicolumn{4}{c}{z$>$1.5} \\ 
\hline
  & mean & $\sigma$ & ${\rm f_{outlier}}$ & NMAD & mean & $\sigma$ & ${\rm f_{outlier}}$ & NMAD & mean & $\sigma$ & ${\rm f_{outlier}}$ & NMAD \\
\hline
All (before template correction) & 0.010 & 0.040 & 6.1\% & 0.031 & 0.013 & 0.042 & 6.9\% & 0.035 & 0.001 & 0.035 & 3.5\% & 0.023 \\
(after template correction)    & -0.002 & 0.037 & 5.5\% & 0.028 & -0.003 & 0.037 & 6.0\% & 0.030 & -0.001 & 0.037 & 3.8\% & 0.023 \\
\hline
H$\leq$24 AB (before template correction) & 0.012 & 0.041 & 5.3\% & 0.032 & 0.013 & 0.042 & 5.8\% & 0.034 & 0.004 & 0.037 & 3.0\% & 0.023 \\
(after template correction)    & -0.002 & 0.037 & 4.8\% & 0.028 & -0.003 & 0.037 & 5.1\% & 0.029 & 0.002 & 0.039 & 3.6\% & 0.024 \\
\hline
H$>$24 AB (before template correction) & -0.001 & 0.039 & 11.7\% & 0.029 & 0.009 & 0.050 & 22.2\% & 0.057 & -0.006 & 0.031 & 4.4\% & 0.022 \\
(after template correction)    & -0.004 & 0.036 & 10.3\% & 0.027 & 0.001 & 0.043 & 19.0\% & 0.046 & -0.008 & 0.030 & 4.4\% & 0.021 \\
\hline
\end{tabular}
\footnote[0]{Note: Outliers are defined as objects with $|\Delta z|/(1+z) > 0.15$, mean and $\sigma$ are calculated after excluding outliers. NMAD is defined as ${\rm 1.48 \times median(|\Delta z|/(1+z))}$.} 
\end{center}
\vspace{-0.5cm}
\end{table*}

Overall, the systematic offset in our catalog is small ($\sim$0.02 mag) for
most UV-to-NIR bands and modest ($\sim$0.05 mag) for ISAAC Ks and IRAC 3.6,
4.5, and 5.8 $\mu$m. IRAC 8.0 $\mu$m suffers from large systematic offsets,
possibly due to PAH emissions. These derived zeropoint offsets depend on
both the redshift distribution of the spec-z training sample and the choice of
the photo-z templates. Therefore, they can only serve as a rough evaluation of
the quality of our catalog. We do not apply these offsets to our catalog.

\subsection{Photometric Redshifts}
\label{quality:photoz}

\begin{figure}[htbp]
\vspace*{-0.5cm}
\center{
\hspace*{-0.5cm}
\vspace*{-1.0cm}
\includegraphics[scale=0.20]{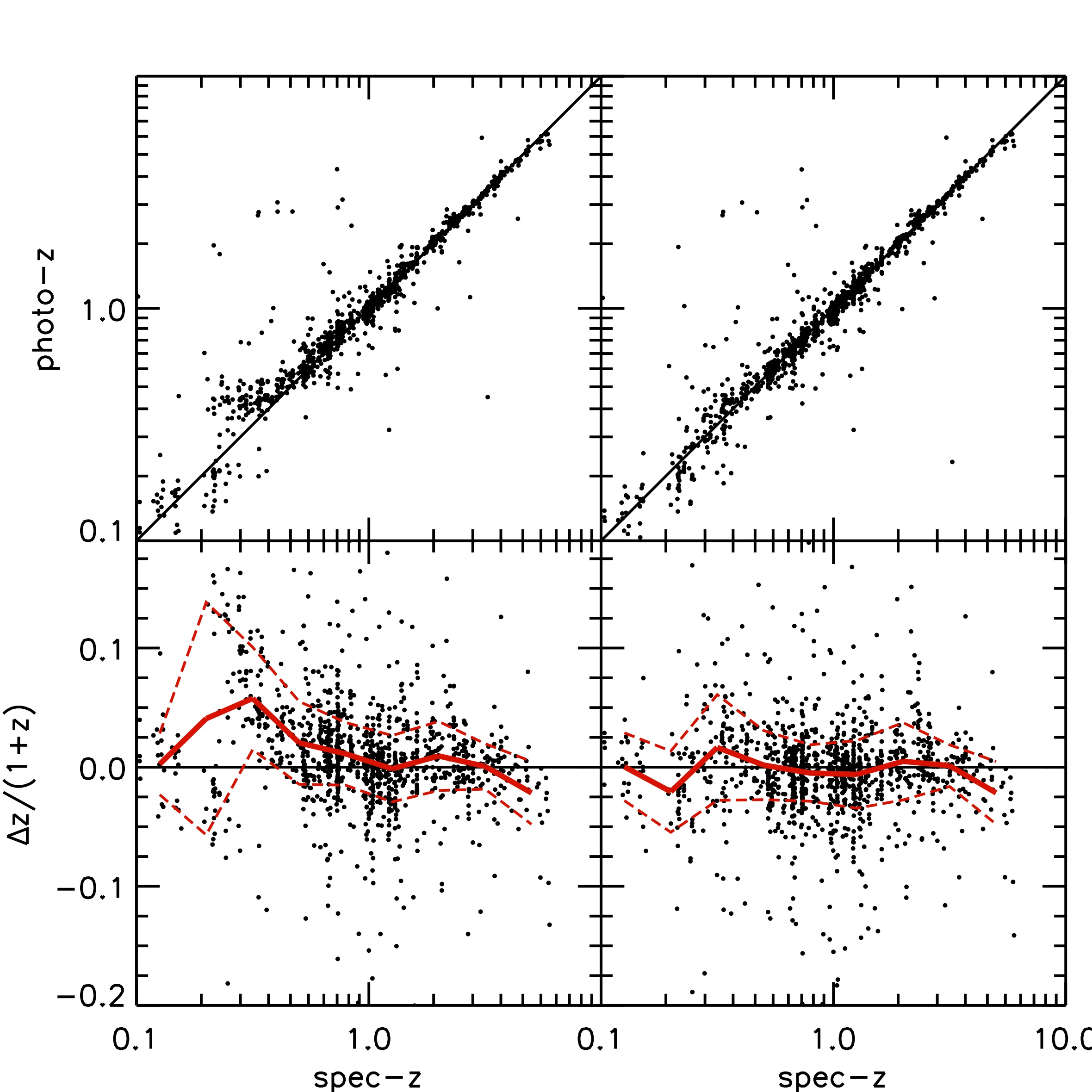}
\vspace*{1.0cm}
\caption[]{Accuracy of photometric redshifts measured with our catalog, {\it
before (left)} and {\it after (right)} applying our template correction.
In each column, the {\it upper} panel shows the comparison of photo-zs
and spec-zs, and the {\it lower} panel shows the relative error of photo-z as a
function of spec-z. The solid and dashed lines in the lower panels stand for
the mean and standard deviation (after 3$\sigma$-clipping) of the sample.
Detailed statistics on the photo-z accuracy are given in Table \ref{tb:photoz}.
\label{fig:photoz}}
}
\vspace{-0.2cm}
\end{figure}

We now measure photometric redshifts (photo-zs) of galaxies in our
spectroscopic sample. We use the PEGASE v2.0 templates, both before and after
applying our template correction to them.  The accuracy of our photo-z
measurement is shown in Figure \ref{fig:photoz}, and detailed statistics on the
accuracy are listed in Table \ref{tb:photoz}.
Our template correction mainly affects objects at z$<$1 and corrects the
systematic overestimate of photo-z among them. After the template correction,
our measurement has the normalized median absolute deviation (NMAD, defined as
${\rm 1.48 \times median(|\Delta z|/(1+z))}$) of 0.028 and an outlier fraction
(defined as $|\Delta z|/(1+z) > 0.15$) of 5.5\% for all galaxies. These values
are slightly better for bright (H$<$24 AB) galaxies, but worse for faint
(H$>$24 AB) objects.

For galaxies at z$>$1.5, the accuracy increases to an NMAD$\sim$0.023 and
outlier fraction of $\sim$4.0\%, regardless of whether template correction is
applied or not. The accuracy also depends little on the brightness of objects.
This high accuracy indicates the importance of the CANDELS NIR data on photo-z
measurement for galaxies at intermediate or high redshift. 
The deep CANDELS photometry in our catalog effectively characterizes the
Balmer/4000 \AA\ break for galaxies at z$>$1.5 and hence yields an accurate
photo-z measurement for them. For low-redshift (z$<$1.5) galaxies, their
photo-zs are determined by both major spectral breaks (Lyman or Balmer) as well
as the Rayleigh--Jeans regime of stellar emission. The latter is quantified by
IRAC photometry in our catalog, which has larger uncertainty than CANDELS NIR
photometry and hence reduces the accuracy of the photo-z of z$<$1.5 galaxies.

It is also possible that the high accuracy of our photo-zs of z$>$1.5
galaxies is due to the narrow range of the spectral types of our spec-z sample,
which are fully covered by our SED-fitting templates. The narrow range of the
spectral types could be physically real because galaxies at z$>$1.5 have
a relatively simple star formation history, or it could be due to the selection
bias of our spectroscopically observed sample. On the other side, our templates
may not fully cover the spectral types of z$<$1.5 galaxies because of their
complicated star formation histories as well as variation in the dust
extinction law, resulting in a decreased photo-z accuracy. In order to improve
the photo-z at z$<$1.5, the uncertainties of SED-fitting templates in stellar
evolutionary track, star formation history, dust extinction law, and spectral
features should be taken into account. For example, \citet{brammer08}
introduced a template error function to incorporate the above uncertainties
together with the photometric uncertainty into the template fitting algorithm.
The template error function minimized systematic errors in their photo-zs.

Overall, our photo-z measurement shows that our photometric catalog is
sufficient to provide accurate photo-zs for galaxies, especially for those at
z$>$1.5, where CANDELS NIR data greatly improve the quality of photo-zs. The
CANDELS team is working on generating a photo-z catalog for the GOODS-S field
(Dahlen et al., 2013, in preparation). 

\section{Application of the Catalog to Studying Galaxies at z=2--4}
\label{app}


\begin{figure}[htbp]
\center{
\hspace*{-0.5cm}
\vspace*{-0.5cm}
\includegraphics[scale=0.6]{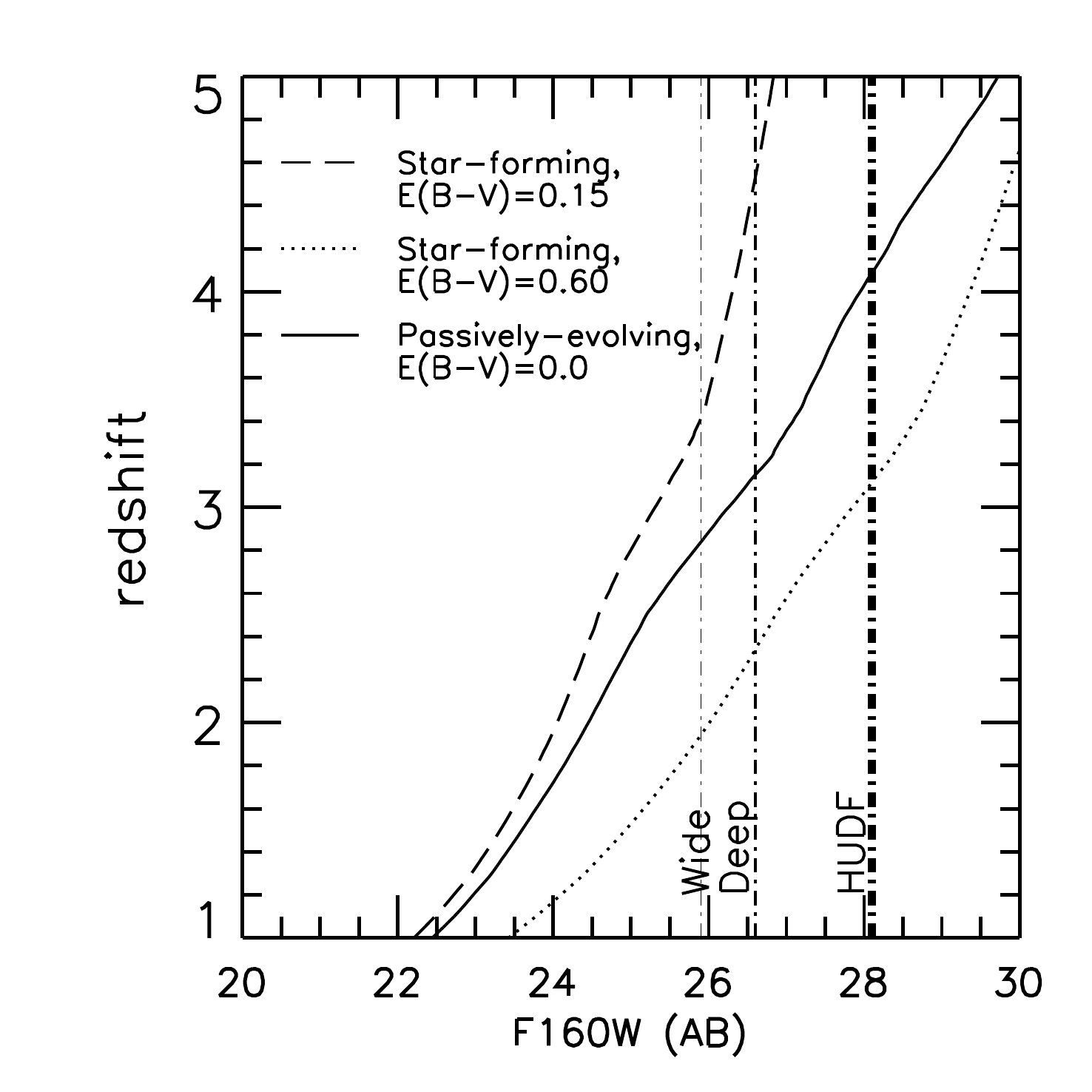}
\caption[]{Detection ability of the CANDELS/GOODS-S catalog. Each curve
shows the F160W magnitude of a model stellar population at different redshift.
All models have stellar mass of ${\rm 10^{10} M_\odot}$ and age of 1 Gyr at the
epoch of the observation. The solid curve denotes passively evolving galaxies
that had a single burst of star formation and has no dust reddening. Dashed and
dotted curves denote populations with a constant star formation history and
different amounts of dust reddening characterized by E(B-V) as indicated.
Vertical lines show the representative 50\% completeness levels in the three
regions of the CANDELS/GOODS-S field.
\label{fig:masscomplete}}
}
\vspace{-0.2cm}
\end{figure}

\subsection{Detection Ability}
\label{app:ability}

Because our catalog is constructed based on F160W detection, the depth and
completeness of our F160W band place the primary limit on our detection
ability. We use galaxy templates drawn from the updated version of
\citet[][CB09]{bc03} to predict the F160W photometry for different types of
galaxies. The detection ability of our catalog is plotted in Figure
\ref{fig:masscomplete}. For simplicity, we only discuss galaxies with stellar
mass of ${\rm 10^{10} M_\odot}$ and age of 1 Gyr at the epoch of observation.
Galaxies with a constant star formation history and low dust extinction
(e.g., E(B-V)=0.15) can be detected with a completeness of 50\% to z$\sim$3.4
and z$\sim$4.6 in our wide and deep regions, respectively. Due to their very
red rest-frame UV and optical colors, star-forming galaxies with high dust-extinction can only
be detected with the same completeness to lower redshift. Galaxies with
E(B-V)=0.6 can only be detected to the 50\% completeness to z$\sim$2.0, 2.4 and
3.0 for the wide, deep, and HUDF regions, respectively. For passively-evolving
galaxies (PEGs), we can detect them with 50\% completeness to z$\sim$2.8, 3.2,
and 4.2 in the wide, deep, and HUDF regions, if they are unreddened.
For galaxies with age older than 1 Gyr, the maximum detectable redshift is
lower. For example, PEGs with age of 2 Gyr can be detected with a 50\%
completeness to z=2.4 and 2.8 for the wide and deep regions. 


\begin{figure}[htbp]
\center{
\vspace*{-0.5cm}
\includegraphics[scale=0.6]{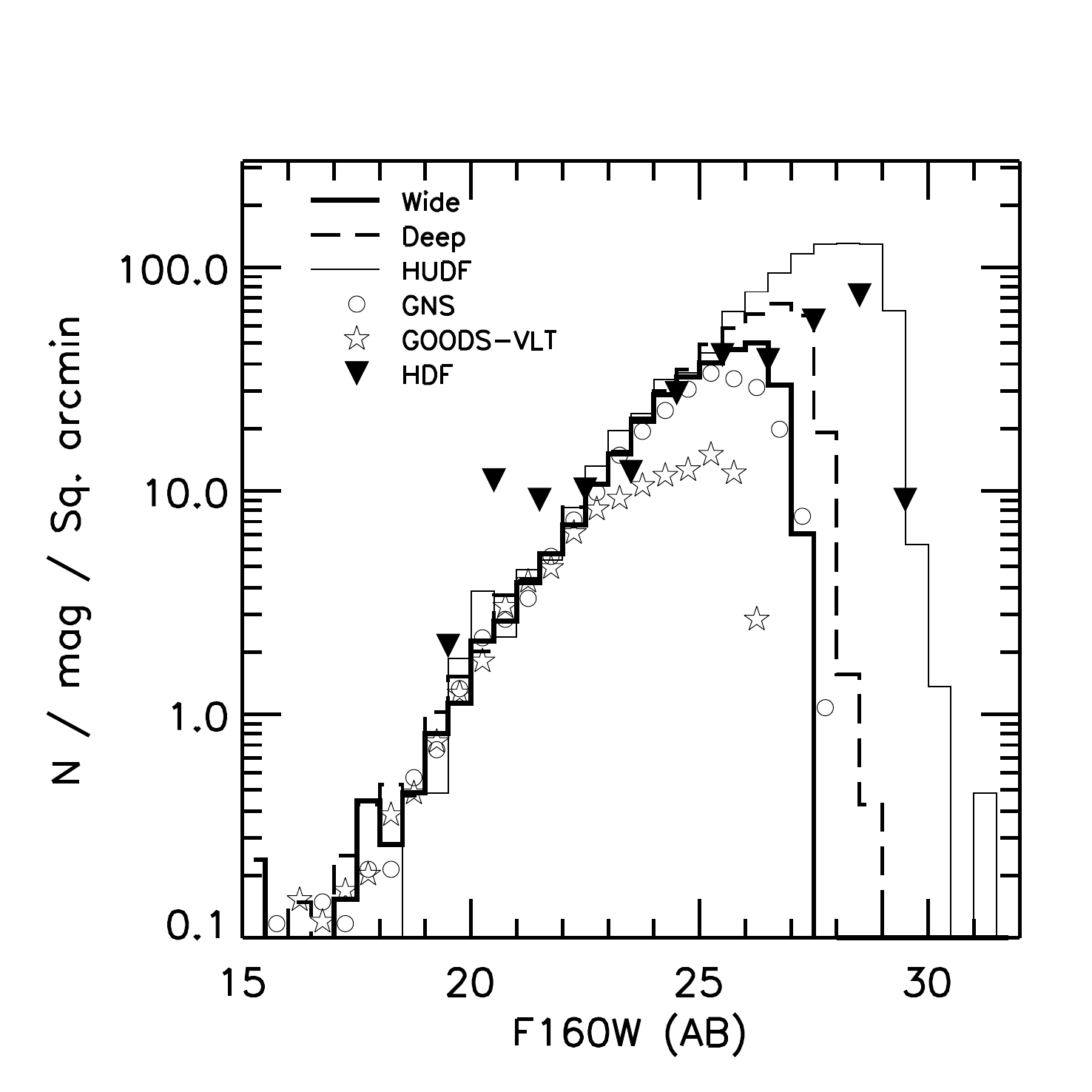}
\caption[]{Differential number count comparison of H-band surveys. Lines show
the three regions of our catalog. Open circles show GOODS NICMOS Survey
\citep[GNS,][]{conselice11}, stars show GOODS-VLT \citep{retzlaff10}, and
triangles show NICMOS HDF Survey \citep{dickinson00,thompson03}. 
\label{fig:ndcomp}}
}
\end{figure}

\begin{figure}[htbp]
\center{
\vspace*{-0.5cm}
\includegraphics[scale=0.35]{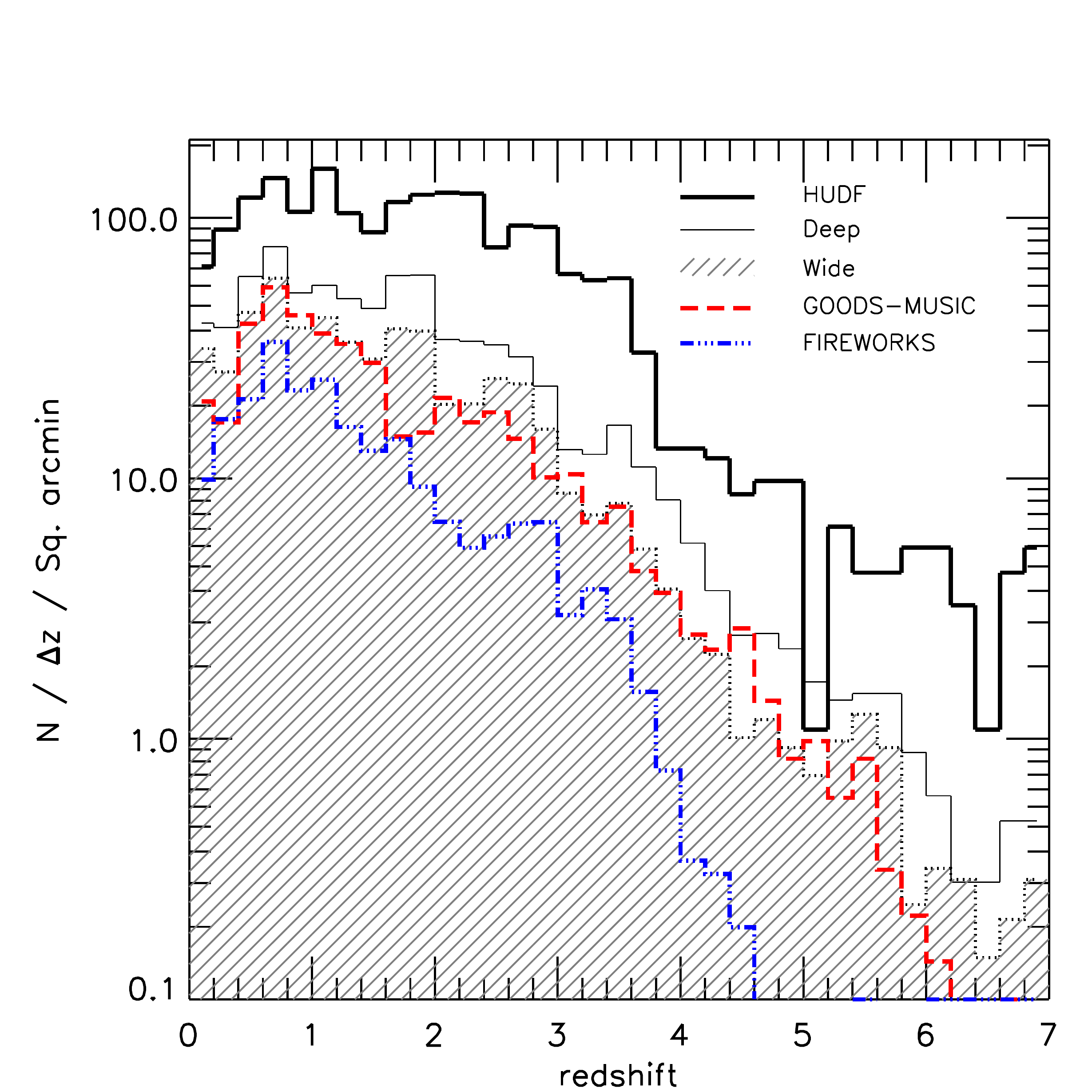}
\caption[]{Photometric redshift distributions of the three regions of the
CANDELS/GOODS-S catalog, GOODS-MUSIC, and FIREWORKS. Each catalog/region is cut
to its 50\% completeness threshold determined by the best-fit power law method
(see Sec. \ref{hstphoto:source}): F160W=25.9, 26.6, 28.1 AB for CANDELS wide,
deep, and HUDF region, F850LP=26.4 AB for GOODS-MUSIC, and Ks=24.1 AB for
FIREWORKS.
\label{fig:photozdist}}
}
\end{figure}

Figure \ref{fig:ndcomp} compares the differential number density of our catalog
with that of other surveys that contain H-band data: GOODS NICMOS Survey
\citep[GNS,][]{conselice11}, GOODS-VLT Survey \citep{retzlaff10}, and NICMOS
HDF Survey \citep{dickinson00,thompson03}. The representative 50\% completeness
limits of GNS and GOODS-VLT estimated through the best-fit power law are 25.1
and 24.3 AB mag, respectively. Our wide (deep) region is 0.8 (1.5) and 1.6
(2.3) mag deeper than GNS and GOODS-VLT. Therefore, our catalog in the
wide (deep) region is able to detect galaxies with stellar mass lower than the
minimum detectable stellar mass of GNS and GOODS-VLT by a factor of two (four)
and four (eight).
The NICMOS HDF Survey is deeper than our deep region but fainter than our HUDF
region. However, its small sky coverage, about 1 arcmin${\rm ^2}$, brings large
cosmic variance and small number statistics, as can be inferred from its
irregular differential number density distribution. 

\begin{figure*}[htbp]
\center{
\hspace*{-1.5cm}
\vspace*{-0.2cm}
\includegraphics[scale=0.45]{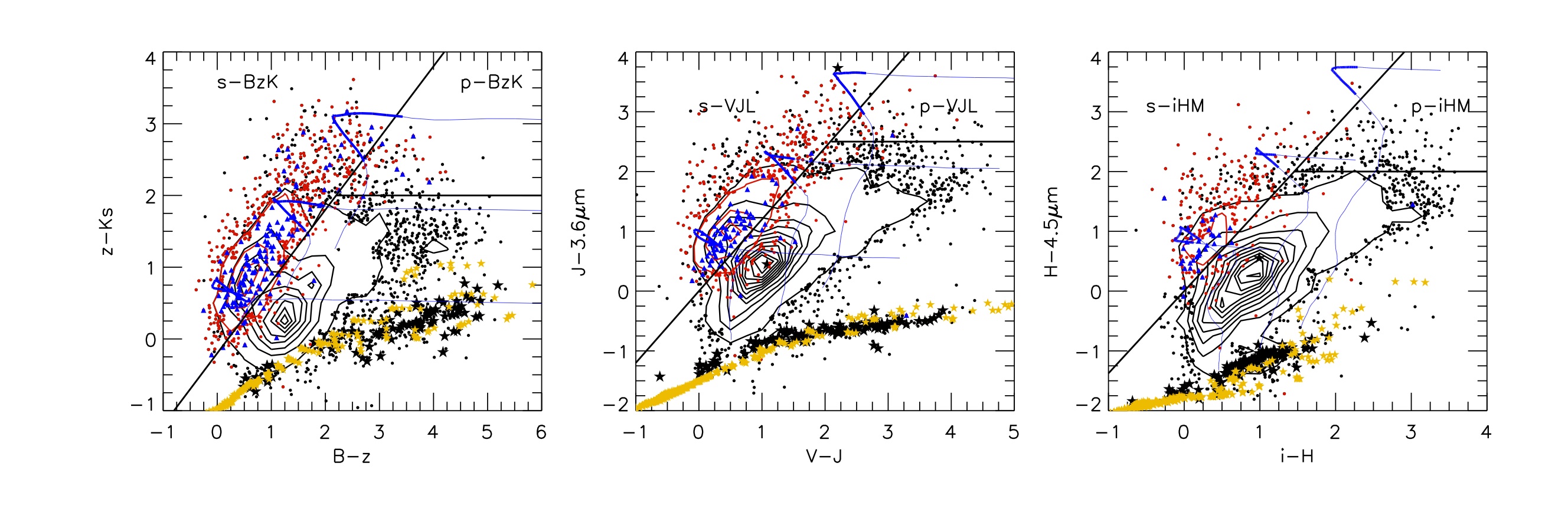}
\caption[]{Color--color criteria of using the Balmer break to select galaxies
at z$\sim$2 (BzK, left), z$\sim$3 (VJL, middle), and z$\sim$4 (iHM, right).
Windows for selecting star-forming (s-) and passively evolving galaxies (p-)
are separated by black solid lines and labeled. In each panel, black contours
and dots show the distribution of all galaxies in our catalog with high S/Ns
($>$5 for z and Ks in BzK, $>$10 for J and 3.6 $\mu$m in VJL, and $>$10 for H
and 4.5 $\mu$m in iHM). Red contours and dots show the distribution of galaxies
with photo-zs within the redshift range of interest ([1.5, 2.5] for BzK, [2.5,
3.5] for VJL, and [3.5, 4.5] for iHM). Blue triangles are galaxies with spec-zs
within the redshift range of interest of each criterion. Blue curves show the
track of star-forming galaxies with constant star-formation history, age of 1
Gyr, and reddening E(B-V)=0.0, 0.3, and 0.6 (for top to bottom in each panel).
In each curve, the thick region shows the redshift range of interest. Black
stars show objects with the F160W SExtractor parameter CLASS\_STAR$>$0.98. 
Yellow stars show stars in the synthetic library of \citet{lejeune97}.
\label{fig:bzkselect}}
}
\end{figure*}

\begin{figure*}[htbp]
\center{
\hspace*{-1.0cm}
\vspace*{-0.2cm}
\includegraphics[scale=0.4]{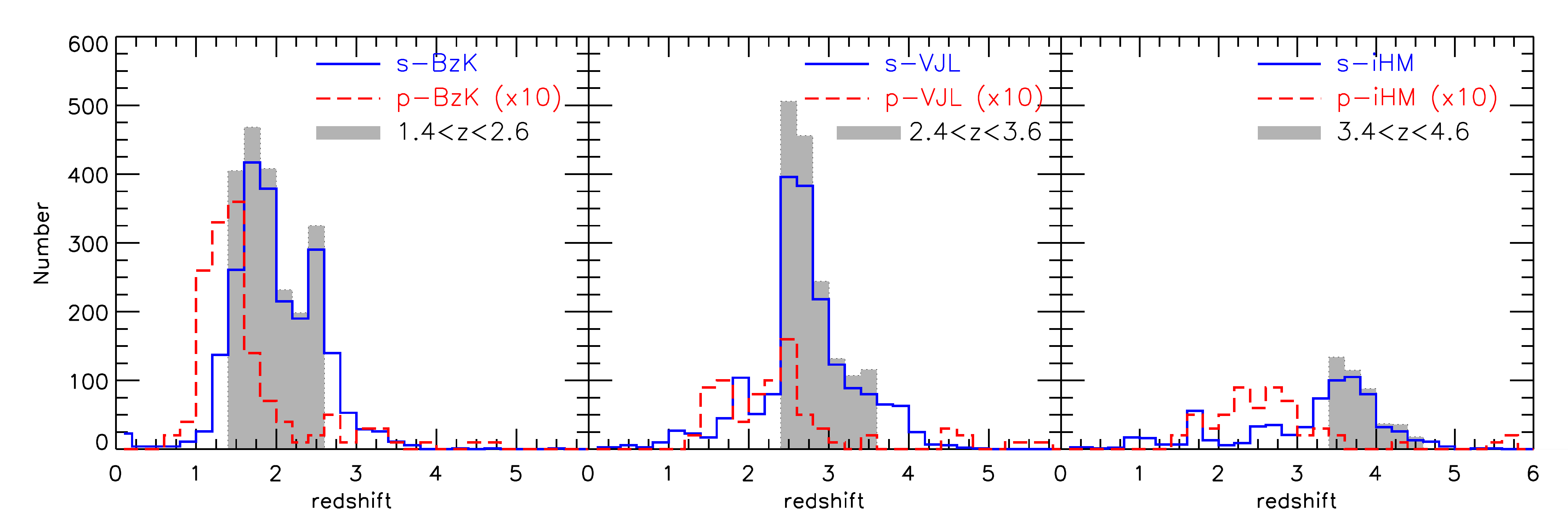}
\caption[]{Photometric redshift distributions of the Balmer break selected galaxies at
z$\sim$2 (left), z$\sim$3 (middle), and z$\sim$4 (right). In each panel, the 
solid blue and dashed red histograms show the distributions of select star-forming (s-) and
passive (p-) galaxies. The histograms of passive galaxies are scaled up by a
factor of 10. The gray histograms show the distribution of all galaxies within the
redshift range of interest of each panel. All histograms include only galaxies
with high S/N: $>$5 for z and Ks in the left panel, $>$10 for J and 3.6 $\mu$m
middle, and $>$10 for H and 4.5 $\mu$m right.
\label{fig:bzkphotoz}}
}
\end{figure*}

To further demonstrate the detection ability of our catalog, we compare
the overall photo-z distributions of the three regions of our catalog to that
of GOODS-MUSIC and FIREWORKS in Figure \ref{fig:photozdist}. We cut each
region/catalog to its 50\% completeness magnitude through the best-fit power
law to the differential source number density of each region/catalog, as
discussed in Sec. \ref{hstphoto:source}.  This threshold magnitude is
F850LP=26.4 AB mag for GOODS-MUSIC and Ks=24.1 for FIREWORKS. 

The photo-z distributions of our wide region and GOODS-MUSIC are similar,
suggesting a comparable detection ability at all redshifts, although our wide
region has a marginal excess of detection ability at z$\sim$2 and z$>$5. Our
deep and HUDF regions surpass the GOODS-MUSIC catalog at all redshifts. The
source number density in our deep (HUDF) region is higher than that of
GOODS-MUSIC by a factor of $\sim$2 ($\sim$6) at 0$<$z$<$5, and by a factor of
$>4$ ($>10$) at z$>$5.  All our three regions have much higher source number
density in each redshift bin than FIREWORKS.

\subsection{Balmer Break Selected Galaxies at z$\sim$2--4}
\label{app:balmer}

Our deep NIR photometry provides an accurate description of the Balmer/4000
\AA\ break for galaxies at z$\sim$2--4, and enables selecting
high-redshift galaxies through broad-band colors that cover the break.
Although current photo-z measurements can achieve a fairly high accuracy, color
selection methods still play an important role in selecting high-redshift
galaxies. The bias, efficiency, and contamination of color selections can be
explicitly determined through Monte Carlo simulation of selecting fake galaxies
with various magnitude, size, color, and star-formation history. Color
selections are also fast and easy to reproduce.

\begin{figure*}[htbp]
\center{
\vspace*{-0.2cm}
\includegraphics[scale=0.5]{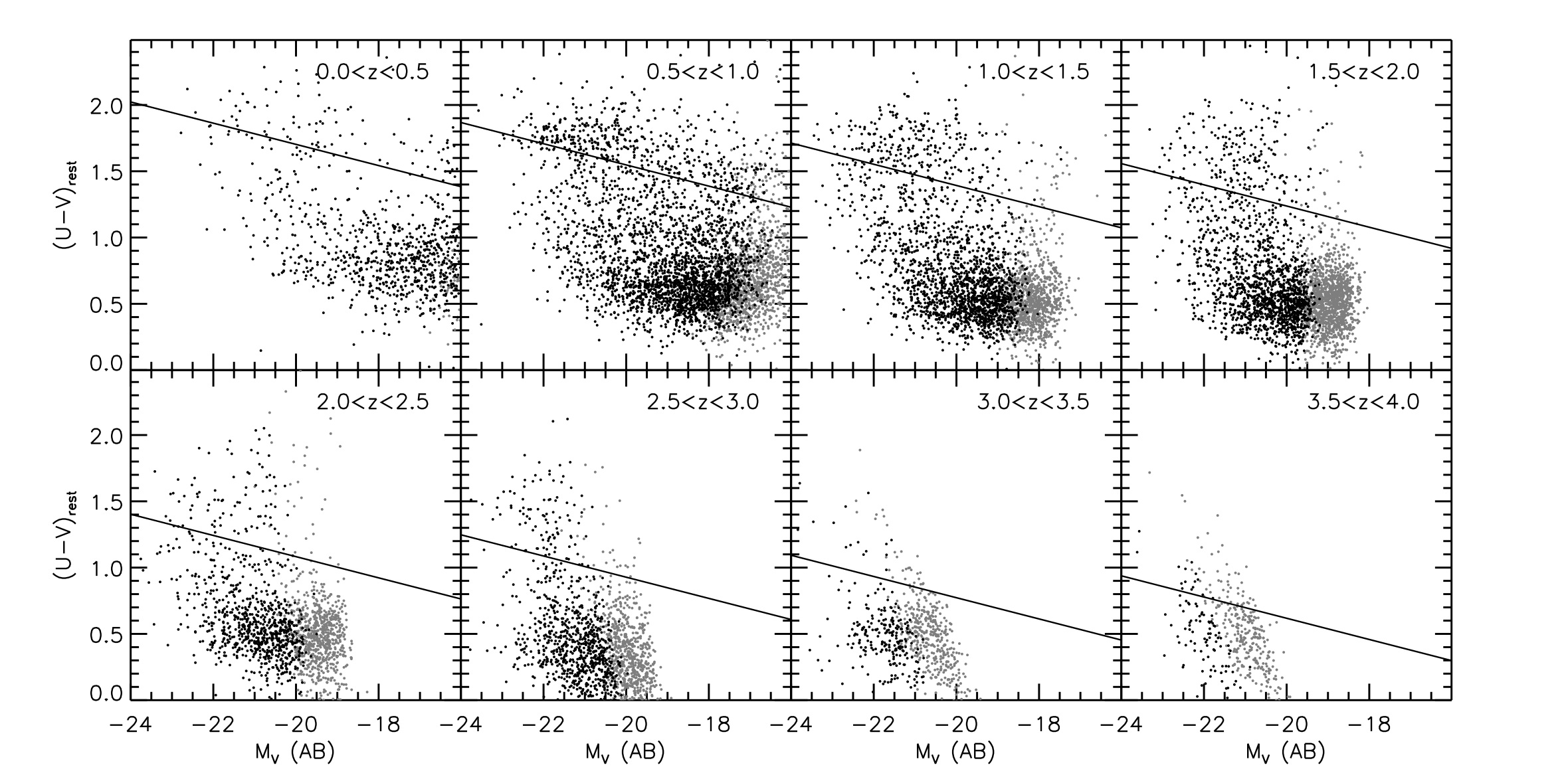}
\caption[]{Rest-frame color--magnitude diagram of galaxies in the
CANDELS/GOODS-S catalog. Black points are galaxies with F160W$<$24.9 AB (50\%
completeness level of the wide region), and gray points are galaxies with
24.9$<$F160W$<$25.9 AB (50\% completeness level of the deep region). The black
line in each panel is the extrapolation of the red--blue separation of
\citet{bell04} to the redshift of the panel.
\label{fig:bimodal}}
}
\end{figure*}

A popular color-color selection criterion using the Balmer break is the BzK
method proposed by \citet{daddi04bzk}. This method, using the (B-z) and (z-K)
colors to quantify the strength and curvature of the Balmer break, is proven
successful at selecting star-forming galaxies independent of their dust
reddening as well as passively evolving galaxies at z$\sim$2. However, due
to the lack of deep NIR photometry of large fields, extending this method to
higher redshifts has not been extensively discussed. It was only extended to
galaxies at z$\sim$3 in the ERS field by \citet{ycguo12vjl} using the (V-J) and
(J-3.6$\mu$m) colors. Our catalog with the deepest NIR photometry in the
GOODS-S field is the first one that makes this Balmer break selection method
practical in large fields. We apply the two selection criteria to our catalog
and also extend them to z$\sim$4.

The selection criteria used here are as follows.
\begin{itemize}
\item[1] Star-forming BzK: $(z-Ks) > (B-z)-0.2$, and passively evolving BzK: $(z-Ks) \leq (B-z)-0.2 \land (z-Ks)>2.0$, where $B$ and $z$ are F435W and F850LP;
\item[2] Star-forming VJL: $(J-L) > 1.2\times(V-J)$, and passively evolving VJL: $(J-L) \leq 1.2\times(V-J) \land (J-L)>2.5$, where $V$, $J$, and $L$ are F606W, F125W, and IRAC 3.6 $\mu$m;
\item[3] Star-forming iHM: $(H-M) > 1.375\times(i-H)$, and passively evolving iHM: $(H-M) \leq 1.375\times(i-H) \land (H-M)>2.0$, where $i$, $H$, and $M$ are F775W, F160W, and IRAC 4.5 $\mu$m.
\end{itemize}
For BzK, we demand S/N$>$5 in the z- and Ks-band. We use the ISAAC Ks-band here
because it covers the entire GOODS-S field. Sommariva et al. (2013, in
preparation) are studying BzK selected sources in both GOODS-S and UDS based on
the HAWK-I Ks-band. For VJL and iHM, we demand S/N$>$10 in the two reddest
bands used in each criterion.

The color--color diagram of each selection criterion is shown in Figure
\ref{fig:bzkselect}. The target redshift range of each criterion is
z$\sim$1.5--2.5 for BzK, 2.5--3.5 for VJL, and 3.5--4.5 for iHM. Galaxies with
photo-zs within each target range are over-plotted in each diagram. These
selection criteria do a fairly good job at selecting galaxies in their own
target redshift range. Figure \ref{fig:bzkphotoz} shows the redshift
distribution of selected galaxies in each criterion. We can roughly estimate
the effectiveness and contamination of each criterion by comparing the redshift
distribution of selected galaxies (blue histogram) and that of total galaxies
(black histogram). The effectiveness, defined as the fraction of galaxies in
the target redshift range that are selected by a given criterion, is about
86\%, 83\%, and 83\% for BzK, VJL, and iHM, respectively. The contamination,
defined as the fraction of selected galaxies that are not within the target
redshift range of the selection criterion, is about 21\%, 30\%, and 50\% for
BzK, VJL, and iHM, respectively. 

We also compare the observed and synthetic stellar locus in the color--color
diagrams to examine our NIR photometry. Since our tests on the ACS--WFC3 colors
of stars show no offset between the observed and synthetic stars (Sec.
\ref{quality:star}), any deviation between the observed and synthetic stars in
these diagrams is caused by the systematic offset of the NIR bands. In the BzK
diagram, the colors of observed stars show good agreement with those of stars
in the synthetic library of \citet{lejeune97}, suggesting no systematic offset
in our ISAAC Ks-band photometry. In the VJL and iHM diagrams, a mild deviation
about $\sim$0.05 mag is found between the observed and synthetic stars. This
deviation may indicate that our 3.6 and 4.5 $\mu$m fluxes are overestimated by
$\sim$0.05 mag. It can also be due to, however, the inaccurate calibration of
the NIR regime of the stellar libraries. 


Passively evolving galaxies at z$>$3 potentially hold important clues to the
processes that quench star-formation in massive galaxies, and to the emergence
of the Hubble sequence.
As discussed above, our catalog is able to detect such galaxies to z$>$3.
Although the redshift distribution of selected passive galaxies (red histograms
in Figure \ref{fig:bzkphotoz}) declines dramatically at z$>$3, we still find 26
galaxies from the passive selection windows. The major contamination source in
the passive selection windows is dusty star-forming galaxies at lower redshift.
In order to exclude the contamination, we remove 13 sources with MIPS 24
$\mu$m detection. If the remaining 13 galaxies are confirmed to have the
passive nature, their star formation activity should have ceased at least
0.5--1 Gyrs ago, i.e., at z$\sim$5. 

\subsection{Red Galaxies at z$\sim$2--4}
\label{app:red}

Based on their positions in the color--magnitude diagram (CMD), local
galaxies can be roughly divided into two populations: blue cloud and
red-sequence \citep[e.g.,][]{strateva01,blanton03b,baldry04}. Galaxies in the
blue cloud are reliably classified as spiral galaxies with ongoing
star-formation, and their color is strongly related to the recent
star-formation history. Galaxies in the red-sequence (except dust-reddened
spirials) generally have little recent star-formation. This color bimodality
has been reported up to z$\sim$3 \citep[e.g.,][]{bell04, cassata08, kriek09,
brammer09, whitaker11}. It is unclear, however, at which cosmic epoch this
color bimodality emerges, owing to limitations of the accuracy of photo-zs and
rest-frame UV--optical colors. 

\begin{figure}[htbp]
\center{
\vspace*{-0.5cm}
\includegraphics[scale=0.22]{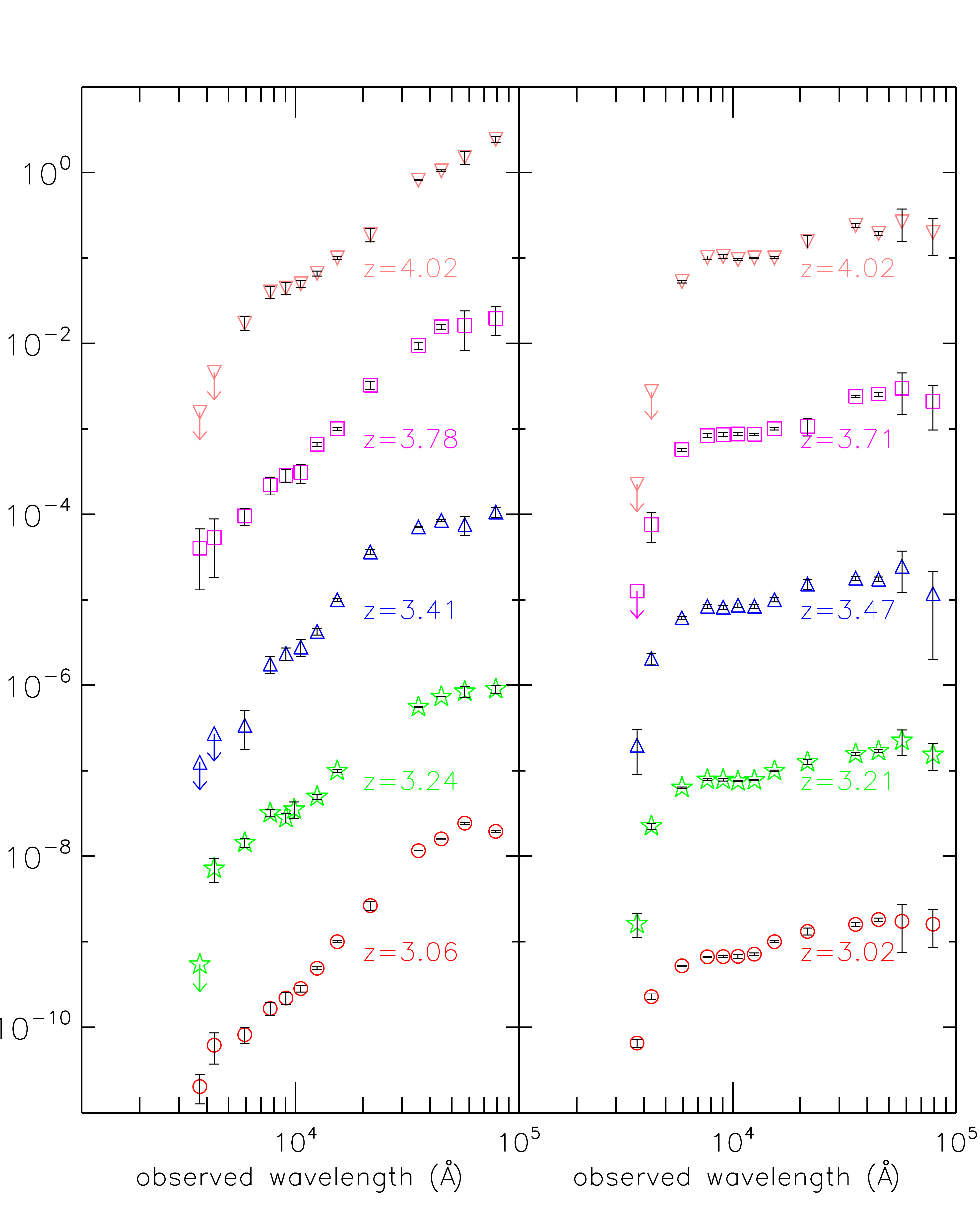}
\caption[]{Examples of SEDs of red (left column) and blue (right column)
galaxies at 3$<$z$<$4 in the CANDELS/GOODS-S catalog. The redshifts of the 
galaxies increase from the bottom to the top, as shown by labels.
\label{fig:sedsample}}
}
\end{figure}

The deep NIR photometry of our catalog provides a unique tool to detect red
galaxies at z$>$3 to investigate the origin of the color bimodality. Figure
\ref{fig:bimodal} shows the rest-frame color--magnitude diagram of galaxies at
different redshifts in our catalog. We use the extrapolation of the red--blue
separation line of \citet{bell04} to separate the blue cloud and red-sequence.
The red-sequence is clearly shown at z$<$2. At 2$<$z$<$3, the red-sequence
becomes less obvious, but there is still a large amount of galaxies which are
well separated from the blue cloud and enter the red region. At z$>$3, due to
the decline of number of galaxies, the separation of the two populations is not
as clear as at lower redshifts. However, a handful of galaxies still have red
enough rest-frame UV--optical color to enter the red region, suggesting the
existence of passive galaxies at z$>$3.

We present a few examples of red and blue SEDs at 3$<$z$<$4 in our catalog in
Figure \ref{fig:sedsample} to show the quality of our photometry. The small
uncertainties on the rest-frame UV--optical colors enable reliable separation
of different types of SEDs of galaxies up to z$\sim$4. However, although red
galaxies at z$\sim$4 suggest the existence of evolved galaxies with little
ongoing star formation at such high redshift, they could also be dusty
star-forming galaxies whose rest-frame UV--optical colors are reddened by dust.
To investigate the nature, stellar masses, and number densities of these red
galaxies is a subject for future works. 

\begin{table*}[htbp]
\begin{center}
\caption{Column Description of the CANDELS GOODS-S Catalog \label{tb:column}}
\begin{tabular}{lll}
\hline\hline
Column No. & Column Title & Description \\
\hline
1 & ID & Object identifier, beginning from 1 \\
2 & IAU Name & \\
3,4 & RA, DEC & Right ascension and declination (J2000.0; decimal degrees) \\

5 & F160W Limiting Magnitude & Limiting magnitude at the
position of the source in the F160W image\footnote[1]{The limiting
magnitude here is derived as $m_{lim} = -2.5 \times log_{10}(\sqrt{A<\sigma^2>})+zp$, where $<\sigma^2>$ is the average of the squared RMS in the SExtractor F160W
segmentation map of each source, $A$ is the area of 1 ${\rm arcsec^2}$, and 
$zp$ is the zero-point of F160W.} \\

6 & FLAGS & SExtractor F160W flag used to designate suspicious sources that
fall in contaminated regions\footnote[2]{A non-contaminated source has a
flag of ``0''. Sources detected on star spikes, halos, and the bright stars
that produce those spikes and halos have a flag of ``1''. Sources detected by
SExtractor at the image edges or on the few artifacts of the F160W image are
assigned a flag of ``2''. Sources with both the flag of ``1'' and ``2'' have a
flag of ``3''.} \\
7 & CLASS\_STAR & SExtractor CLASS\_STAR parameter in the F160W band \\
8--58 & Flux, Flux\_Err, Weight & Triplet of flux, flux uncertainty, and weight in each filter. Sources that are not observed have (-99.00, -99.00, 0). \\
      &                        & In \hst\ bands, the weight is the exposure time of the source, while in other bands, it is a relative weight. \\
      &                        & Filters are included in order: CTIO\_U, VIMOS\_U, F435W, F606W, F775W, F814W, F850LP, F098M, F105W, \\
      &                        & F125W, F160W, ISAAC\_Ks, HAWK-I\_Ks, 3.6 $\mu$m, 4.5 $\mu$m, 5.8 $\mu$m, and 8.0 $\mu$m \\
59, 60 & FLUX\_ISO, FLUXERR\_ISO & F160 isophotal flux and
flux error\footnote[3]{For sources whose isophotal radius smaller than 2.08
pixels, these parameters are replaced by FLUX\_APER and FLUXERR\_APER measured
within a radius of 0\farcs125 (2.08 pixels). See Sec. \ref{hstphoto:others} for
details.} \\
61, 62 & FLUX\_AUTO, FLUXERR\_AUTO & F160 AUTO flux and flux error\\
63 & FWHM\_IMAGE & FWHM of the F160W image of object, in unit of pixel (1 pixel = 0\farcs06) \\
64, 65 & A\_IMAGE, B\_IMAGE & F160W profile RMS along major and minor axis (pixel) \\
66 & KRON\_RADIUS & F160W band Kron radius from SExtractor (in unit of A\_IMAGE or B\_IMAGE) \\
67, 68, 69 & FLUX\_RADIUS & F160W band FLUX\_RADIUS with the fraction of light of 0.2, 0.5, and 0.8 from SExtractor (pixel) \\
70 & THETA\_IMAGE & F160W position angle (degree) \\
71 & Apcorr & Ratio of SExtractor FLUX\_AUTO and FLUX\_ISO in the F160W band \\
72 & HOT\_FLAG & A flag to designate if the source enters the catalog as detected in the cold mode (=0) or in the hot mode (=1) \\
73 & ISOAREAF\_IMAGE & SExtractor F160W Isophotal area (filtered) above Detection threshold (${\rm pixel^2}$) \\
\hline
\end{tabular}
\end{center}
\vspace{-0.3cm}
\end{table*}

\section{Summary}
\label{summary}

We present a UV-to-mid infrared multi-wavelength catalog in the CANDELS/GOODS-S
region, combining the newly obtained CANDELS \hst/WFC3 F105W, F125W, and F160W
data with existing public data. The catalog is based on source detection in the
WFC3 F160W band. To maximize the scientific yields of our catalog, we construct
a ``max-depth'' F160W mosaic, which includes data from the CANDELS deep and
wide observations as well as two previous \hst/WFC3 programs, ERS and HUDF09.
The F160W mosaic, providing the deepest photometry for each individual source
in GOODS-S, reaches a 5$\sigma$ limiting depth (within an aperture of
0\farcs17, the FWHM of the F160W) of 27.4, 28.2, and 29.7 AB mag for CANDELS
wide, deep + ERS, and HUDF regions, respectively. The representative 50\%
completeness of our catalog reaches 25.9, 26.6, and 28.1 AB mag in the F160W
band for the three regions. The catalog contains 34930 sources in total.

In addition to WFC3 NIR bands, the catalog also includes data from UV (U-band
from both CTIO/MOSAIC and VLT/VIMOS), optical (GOODS \hst/ACS F435W, F606W,
F775W, F814W, and F850LP), and IR (ERS WFC3 F098M, VLT/ISAAC Ks, VLT/HAWK-I Ks,
and Spitzer/IRAC 3.6, 4.5, 5.8, 8.0 $\mu$m) observations. We use SExtractor to
measure photometry for all available \hst\ bands from PSF matched images. For
other low-resolution bands, whose FWHMs of PSFs vary by almost a factor of 10,
we use our profile template fitting package, TFIT, to measure the uniform
photometry among them. 

The quality of the multi-wavelength catalog is thoroughly tested. First, the
comparison between the ACS--WFC3 colors of stars in our catalog and those of
stars in synthetic stellar libraries shows excellent agreements. Second, the
difference of photometry between our catalog and GOODS-MUSIC and FIREWORKS is
nearly zero over the magnitude range to $\sim$24 AB mag, with the worst offset
being $\sim$0.05 mag. We discuss the possible reasons of the deviations between
our and other catalogs at the faint end. Last, the zeropoint offsets measured
by comparing the observed SEDs to the best-fit stellar population synthesis
templates of spectroscopically observed galaxies are $\lesssim$0.02 mag for most
UV-to-NIR bands. The photo-zs measured with our catalog can reach the accuracy
of NMAD$\sim$0.028.

Our catalog is able to detect star-forming galaxies with stellar mass of ${\rm
10^{10} M_\odot}$ to z$\sim$3.4, 4.6, and 8.0 at a 50\% completeness in the
wide, deep, and ultra-deep regions, respectively. Passive galaxies with the
same stellar mass can be detected at a 50\% completeness to z$\sim$2.8, 3.2,
and 4.2 in above regions. Our catalog in the wide and deep regions is about
1--2 mag deeper than other public H-band surveys. The minimum detectable
stellar mass in the wide and deep regions of our catalog is therefore smaller
than that of other H-band surveys by a factor of about 3--6.

To provide an example of applying our catalog to study distant galaxies, we use
color criteria to select both star-forming and passively evolving galaxies at
z=2--4 through the strength and curvature of the Balmer Break. We examine the
redshift distribution of selected galaxies and discuss the effectiveness and
contamination of the selection criteria. The detection of red galaxies with
obvious Balmer break in our catalog suggests the existence of passive galaxies
at z$>$3.

We also study the evolution of the rest-frame color--magnitude diagram,
(U-V) vs. ${\rm M_V}$, from z=0 to z=4 to investigate the emergence of the
color bimodality in the universe. A handful of red galaxies, possibly a
combination of passive galaxies and dusty star-forming galaxies, are detected
in our catalog at z$\sim$4.

%

The CANDELS GOODS-S multi-wavelength catalog and its associated files are
publicly available on the CANDELS website, in the Mikulski Archive for Space
Telescopes (MAST), via the on-line version of the article, and the Centre de
Donnees astronomiques de Strasbourg (CDS) as well as in the Rainbow
Database.\\

We thank the anonymous referee for constructive comments that improve this
article. Support for program number HST-GO-12060 was provided by NASA through
a grant from the Space Telescope Science Institute, which is operated by the
Association of Universities for Research in Astronomy, Incorporated, under NASA
contract NAS5-26555. YG and the authors from UCSC acknowledge support from NASA
HST grant GO-12060.10-A and NSF grant AST-0808133. JSD acknowledges the support
of the European Research Council via the award of an Advanced Grant and the
support of the Royal Society via a Wolfson Research Merit Award. This work is
based in part on observations made with the Spitzer Space Telescope, which is
operated by the Jet Propulsion Laboratory, California Institute of Technology
under a contract with NASA. Support for this work was provided by NASA through
an award issued by JPL/Caltech.

{\it Facilities}: \hst\ (ACS and WFC3) and \spitzer\ (IRAC)




%
%
%

\end{document}